\documentclass[10pt,aps,prd,longbibliography,onecolumn,showpacs,amsmath,amssymb,superscriptaddress,nofootinbib,floatfix,preprintnumbers,secnumarabic]{revtex4-1}

\usepackage{graphicx}
\usepackage[usenames,dvipsnames]{xcolor}
\newcommand{\remove}[1]{}
\usepackage{xspace}
\definecolor{darkblue}{rgb}{0,0,0.5}
\definecolor{darkgreen}{rgb}{0.1,0,0.3}
\definecolor{darkred}{rgb}{0.6,0,0}
\usepackage[bookmarks=false, bookmarksnumbered=true, colorlinks=true, urlcolor=darkblue, citecolor=darkgreen, linkcolor=darkred, breaklinks=true]{hyperref} 
\usepackage{siunitx}
\usepackage[normalem]{ulem}
\usepackage[capitalise]{cleveref}

\usepackage{standalone}
\usepackage{multirow}
\usepackage{adjustbox}
\usepackage{booktabs}
\usepackage{appendix}
\usepackage{commath}
\usepackage{rotating}

\usepackage{url}
\usepackage{braket}
\usepackage{tabularx}
\usepackage{siunitx}
\usepackage[utf8]{inputenc}
\usepackage{xspace}
\usepackage{bm}
\usepackage{braket}
\usepackage[T1]{fontenc}
\usepackage{tikz}
\usepackage[ISO]{diffcoeff}

\renewcommand{\vec}[1]{\bm{#1}}

\usepackage{fullpage}
\usepackage{geometry}      
\newcolumntype{Y}{>{\centering\arraybackslash}X}
\newcolumntype{Z}{>{\centering\arraybackslash}p{0.2\hsize}X}

\newcommand{\cevns}{CE$\nu$NS\xspace}

\newcommand{\ee}{\textsubscript{ee}}

\newcommand\thefont{\expandafter\string\the\font}

\definecolor{lime}{HTML}{A6CE39}
\DeclareRobustCommand{\orcidicon}{%
    \begin{tikzpicture}
    \draw[lime, fill=lime] (0,0) 
    circle [radius=0.16] 
    node[white] {{\fontfamily{qag}\selectfont \tiny ID}};
    \draw[white, fill=white] (-0.0625,0.095) 
    circle [radius=0.007];
    \end{tikzpicture}
    \hspace{-2mm}
}
\newcommand{\orcid}[1]{\href{https://orcid.org/#1}{\orcidicon}}

\begin{document}

\preprint{IFT-UAM/CSIC-23-89}

\vspace*{1cm}
\title{Measuring the Sterile Neutrino Mass in Spallation Source and Direct Detection Experiments}

\author{D. Alonso-Gonz\'alez\orcid{0000-0002-7572-9184}}\email{david.alonsogonzalez@uam.es}
\affiliation{Instituto de F\' \i sica Te\'orica, IFT-UAM/CSIC, 28049 Madrid, Spain}
\affiliation{Departamento de F\' \i sica Te\'orica, Universidad Aut\'onoma de Madrid, 28049 Madrid, Spain}

\author{D.W.P. Amaral\orcid{0000-0002-1414-932X}}
\email{dorian.amaral@rice.edu}
\affiliation{Department of Physics and Astronomy, Rice University, Houston, TX 77005, USA}

\author{A. Bariego-Quintana\orcid{0000-0001-5187-7505}}
\email{adriana.bariego@gmail.com}
\affiliation{Instituto de F\' \i sica Corpuscular (CSIC - Universitat de Val\`encia),  46980 Paterna, Valencia, Spain}

\author{D. Cerde\~no\orcid{0000-0002-7649-1956}}
\email{davidg.cerdeno@uam.es}
\affiliation{Instituto de F\' \i sica Te\'orica, IFT-UAM/CSIC, 28049 Madrid, Spain}
\affiliation{Departamento de F\' \i sica Te\'orica, Universidad Aut\'onoma de Madrid, 28049 Madrid, Spain}

\author{M. de los Rios\orcid{0000-0003-2190-2196}}
\email{martin.delosrios@uam.es}
\affiliation{Instituto de F\' \i sica Te\'orica, IFT-UAM/CSIC, 28049 Madrid, Spain}
\affiliation{Departamento de F\' \i sica Te\'orica, Universidad Aut\'onoma de Madrid, 28049 Madrid, Spain}

\begin{abstract}
    We explore the complementarity of direct detection (DD) and spallation source (SS) experiments for the study of sterile neutrino physics. We focus on the sterile baryonic neutrino model: an extension of the Standard Model that introduces a massive sterile neutrino with couplings to the quark sector via a new gauge boson. In this scenario, the inelastic scattering of an active neutrino with the target material in both DD and SS experiments gives rise to a characteristic nuclear recoil energy spectrum that can allow for the reconstruction of the neutrino mass in the event of a positive detection. We first derive new bounds on this model based on the data from the COHERENT collaboration on CsI and LAr targets, which we find do not yet probe new areas of the parameter space. We then assess how well future SS experiments will be able to measure the sterile neutrino mass and mixings, showing that masses in the range $\sim15-50$~MeV can be reconstructed. We show that there is a degeneracy in the measurement of the sterile neutrino mixing that substantially affects the reconstruction of parameters for masses of the order of 40~MeV. Thanks to their lower energy threshold and sensitivity to the solar tau neutrino flux, DD experiments allow us to partially lift the degeneracy in the sterile neutrino mixings and considerably improve its mass reconstruction down to 9~MeV. Our results demonstrate the excellent complementarity between DD and SS experiments in measuring the sterile neutrino mass and highlight the power of DD experiments in searching for new physics in the neutrino sector.

\end{abstract}

\maketitle

\section{Introduction}

The neutrino sector remains one of the most promising places to look for new physics beyond the Standard Model (SM). Amongst the most obvious open problems, the SM offers no explanation for the origin of neutrino masses. A generic prediction of new physics models for neutrino masses is the presence of new sterile neutrino states, which have very small interactions with the SM ones. The masses of these new exotic states depend on the actual mechanism by which neutrinos acquire a mass, but an interesting range of values is the MeV.

The search for sterile neutrinos involves different types of experimental probes and the constraints depend strongly on the mass range of the new states. For example, sterile neutrinos have been widely searched for in meson decays, where masses of up to hundreds of MeV in peak searches of pion and kaon decays have been probed \cite{Britton:1992xv,Britton:1993cj,Aguilar-Arevalo:2019owf,NA62:2020mcv,T2K:2019jwa,NA62:2020mcv}, and heavier steriles have been searched for in neutrino beam dump experiments \cite{Bergsma:1985is,CooperSarkar:1985nh,Abreu:1996pa,Vaitaitis:1999wq}. In our regime of interest (tens of MeV), bounds can be derived through their possible direct production processes. This could be observed in solar neutrino data \cite{Goldhagen:2021kxe}, atmospheric neutrino data \cite{Dentler:2018sju}, or neutrino beam experiment data \cite{Forero:2021azc} like MINOS/MINOS+ \cite{MINOS:2017cae}. In addition, the presence of an extra sterile neutrino may have a non-negligible impact on different cosmological observations depending on its mass and couplings \cite{Abazajian:2012ys}. For example, long-lived sterile neutrinos with masses of the order of MeV may alter Big Bang nucleosynthesis and the expansion rate of the universe \cite{Sabti:2020yrt, Abdullahi:2022jlv}. Moreover, sterile neutrinos decaying before recombination may affect the cosmic microwave background anisotropies \cite{Vincent:2014rja,Boyarsky:2021yoh}.

Experiments situated at spallation source (SS) facilities have recently become excellent probes of new neutrino physics. Most notably, the COHERENT collaboration \cite{COHERENT:2015mry} has been able to observe, for the first time, a very rare SM phenomenon: the coherent elastic scattering of neutrinos with nuclei (\cevns). The results from both the first run on a CsI target \cite{COHERENT:2017ipa} and a second run that employed LAr in the CENNS-10 detector \cite{COHERENT:2020iec} are compatible with the SM prediction \cite{Freedman:1973yd,Drukier:1984vhf}. This has been used to derive limits on new physics in the neutrino sector (see, for example, Refs.~\cite{Kosmas:2017zbh,Miranda:2020syh,Abdullah:2018ykz,Bauer:2018onh,Miranda:2020zji,Bauer:2020itv}), with particular attention to what future detectors can achieve. Planned experiments include CENNS610 \cite{Akimov:2020pdx} (an extension of CENNS-10 LAr \cite{COHERENT:2019kwz}), CCM \cite{ccm}, and efforts in the European Spallation Source facility \cite{Baxter:2019mcx}. The bounds from COHERENT and the sensitivity of the planned detectors are generally interpreted in models with low-mass mediators (or using an effective description in terms of non-standard neutrino interactions), which alters the SM prediction for \cevns \cite{Miranda:2020tif,Abdullah:2018ykz,Banerjee:2018eaf,Papoulias:2019txv,Khan:2019cvi}. Likewise, they are applicable to inelastic processes that involve the up-scattering to a heavy neutrino state, for example through the presence of a nonzero neutrino transition magnetic moment \cite{Bolton:2021pey, Miranda:2021kre,DeRomeri:2022twg}, or even to a dark fermion \cite{Candela:2023rvt}.

In parallel, underground experiments searching directly for dark matter particles have become increasingly sensitive. Planned detectors, especially those based on liquid noble gases, feature extremely clean, ton-scale targets with excellent background discrimination that will soon enable them to measure \cevns from solar neutrinos. Although this would constitute a serious background for dark matter searches, it also offers the unique possibility to test new neutrino physics \cite{Cerdeno:2016sfi,Dutta:2017nht,Gelmini:2018gqa,Essig:2018tss, Amaral:2020tga, Dutta:2020che, Amaral:2021rzw,Munoz:2021sad,deGouvea:2021ymm} in a way that is complementary to that of dedicated neutrino detectors. The main advantages of these direct detection (DD) experiments are that they can probe both electron and nuclear recoils, which makes them a perfect complement to SS and oscillation experiments \cite{Amaral:2023tbs}, and that they are also sensitive to the tau neutrinos in the solar flux.

The sensitivity of DD experiments to observe heavy neutrino states was studied in Ref.~\cite{Shoemaker:2018vii} for the particular case of the neutrino dipole portal, showing that current xenon-based detectors could significantly improve existing astrophysical bounds. The neutrino dipole portal was considered to account for the apparent excess in the low-energy data from electronic recoils in the XENON1T experiment \cite{XENON:2020rca,Shoemaker:2020kji}. However, this solution was seriously limited by other experimental constraints \cite{Brdar:2020quo}, and the excess was not reproduced by XENONnT \cite{XENON:2022ltv}.  Since the coupling of a sterile neutrino to the leptonic sector is in general severely limited by experimental searches, in this article we will focus on the potential interactions with the quark sector. These are more difficult to probe, but they could lead to changes in the predicted nuclear recoil rates in DD and SS experiments that could be accessible in near future experiments. For concreteness, in this work we set up to study the sterile baryonic neutrino (SBN) \cite{Pospelov:2011ha} as an example of models in which the active neutrinos can up-scatter to heavy states.

More specifically, in this article we study the potential of DD and SS experiments to not only detect the sterile neutrino but also reconstruct its parameters---namely, its mass and mixings with the active neutrinos. Our main goal is to determine the conditions under which the sterile neutrino mass can be unambiguously measured (distinguished from zero).

In \cref{sec:sbn}, we introduce an effective construction based on the sterile baryonic neutrino model and determine the new inelastic contribution to neutrino-nucleus scattering. In \cref{sec:spallation}, we address the prospects for upcoming SS experiments. In \cref{sec:direct}, we extend the analysis to include future xenon-based DD experiments. Finally, in \cref{sec:complementarity}, we study the complementary role of DD and SS experiments. We present our conclusions in \cref{sec:conclusions}.

\section{The Sterile Baryonic Neutrino}
\label{sec:sbn}

We introduce a dark sector consisting of a new vector mediator, $Z'$, stemming from a broken $\mathrm{U}(1)_B$ gauge symmetry and a new \textit{baryonic} sterile neutrino, $\nu_b$, that is also charged under this new symmetry \cite{Pospelov:2011ha}. For the purpose of this work, we regard this model as an effective theory, and we do not address its possible anomaly-free UV completion. The relevant part of our Lagrangian is given by 
\begin{equation} \label{eq:Lagrangian}
    \mathcal{L} \supset \frac{m_{Z'}^{2}}{2}Z'^{\mu} Z'_{\mu} + g_{b} Z'^{\mu} \overline{\nu}_b \gamma_\mu \nu_b  + \frac{1}{3}g_{q} Z'^\mu \sum_{q} \overline{q} \gamma_{\mu} q \, .
\end{equation}
Here, $m_Z'$ is the mass of the new boson, $g_{b}$ is its gauge coupling to the baryonic neutrino and $g_{q}$ to the quarks, and the sum runs over all quark flavours $q$. In this model, a generic flavour eigenstate, $\ket{\nu_\alpha}$, can then be written as a linear combination of mass eigenstates, $\ket{\nu_i}$, as
\begin{equation}
    \ket{\nu_\alpha} = \sum_{i=1}^{4} U_{\alpha i}^* \ket{\nu_i}\,,
    \label{eq:mass-decomp}
\end{equation}
where $\ket{\nu_4}$ is the new mass eigenstate with mass $m_4$, and $\alpha \in \{e,\, \mu,\, \tau,\, b\}$.

From \cref{eq:Lagrangian}, and defining the coupling $g_{Z'} \equiv \sqrt{g_bg_q}$, the neutrino-nucleus up-scattering process $\nu_\alpha A\rightarrow \nu_4A$  has amplitude
\begin{equation}
    \mathcal{M}_{\alpha 4}=\frac{g_{Z'}^2}{q^2-m_{Z'}^2}l^\mu h_\mu\,,
\end{equation}
where $q^2$ is the square-momentum exchange with the nucleus, $h^\mu$ is the nucleus transition amplitude for the nuclear ground state $A$, and $l^\mu$ is the leptonic transition amplitude. Using \cref{eq:mass-decomp} to re-write the dark baryonic current in terms of the mass eigenstates, we have that
\begin{equation}
    \begin{split}
    l^\mu \equiv \bra{\nu_4}\overline{\nu}_b \gamma_\mu \nu_b \ket{\nu_\alpha}&=\sum_{ijk}\bra{\nu_4} U_{\alpha k}^* U_{b i}^* U_{b j} \overline{\nu}_j \gamma_\mu \nu_i\ket{\nu_k} = \sum_i U_{\alpha i}^* U_{b 4}^* U_{b i} \bra{\nu_4} \overline{\nu}_4 \gamma_\mu \nu_i\ket{\nu_i} \\ &\simeq U_{\alpha 4}^* \bra{\nu_4} \overline{\nu}_4 \gamma_\mu \nu_i\ket{\nu_i}\,,
    \end{split}
\end{equation}
where, in the last step, we have assumed that $\left|U_{b i}\right|\ll \left|U_{b 4}\right|$ for $i\not=4$ and that $ \left|U_{b4}\right|^2 \simeq 1$ \cite{Arguelles:2022lzs}. The differential neutrino-nucleus up-scattering cross section then follows:
\begin{equation}
    \diff{\sigma_{\alpha 4}}{E_R}=\frac{g_{Z'}^4A^2\left|U_{\alpha4}\right|^2m_A}{2\pi E_\nu^2{\left(2m_AE_R+m_{Z'}^2\right)}^2}\left[4E_\nu^2-2E_R\left(m_A-E_R+2E_\nu\right)-\frac{m_4^2}{m_A}\left(m_A-E_R-E_\nu\right)\right]F^2(E_R),\,
    \label{eq:sigma_a4}
\end{equation}
where $m_A$ is the mass of the target nucleus, $E_\nu$ is the energy of the incoming neutrino, and $E_R$ is the nuclear recoil energy. For the nuclear form factor $F^2(E_R)$, which arises from the hadronic part of the amplitude, we use the Helm form factor \cite{Helm:1956zz} with the parametrisation introduced in Ref.~\cite{Lewin:1995rx}. This new {\em inelastic} scattering process provides an extra contribution to the usual SM {\em elastic} neutrino-nucleus scattering, which takes place through \cevns and has the following differential cross section,
\begin{equation}
    \diff{\sigma_{{\rm CE}\nu{\rm NS}}}
    {E_R}=\frac{G_F^2}{4\pi}Q_\nu^2m_A\left(1-\frac{m_AE_R}{2E_\nu^2}\right)F^2(E_R)\ ,
    \label{eq:sigma_cevns}
\end{equation}
where $G_F$ is the Fermi constant, and $Q_\nu \equiv N-\left(1-4\sin^2\theta_W\right)Z$ is the SM coherence factor in terms of the Weinberg angle, $\theta_W$, and the number of neutrons, $N$, and protons, $Z$.

Note that, for the characteristic recoil energies at SS experiments ($E_R \lesssim \SI{100}{\kilo\electronvolt}$) and DD experiments ($E_R \lesssim \SI{10}{\kilo\electronvolt}$), the cross section in \cref{eq:sigma_a4} can be interpreted as being proportional to the effective coupling $g_{Z'}^4 |U_{\alpha 4}|^2 / m_{Z'}^4$. As both of these types of experiments are sensitive to this product of model parameters, they are only able to make inferences on this effective coupling. Since the focus of our analysis is the physics underlying the baryonic neutrino, we choose to fix the parameters related to the new vector mediator to $m_{Z'} = \SI{1}{\giga\electronvolt}$ and $g_{Z'} = 4 \times 10^{-3}$, taking into account the constraints found in Ref.~\cite{Foguel:2022ppx}. Thus, without loss of generality, for as long as $m_{Z'}^2$ remains greater than the momentum transfer at these experiments, our results can simply be rescaled by the factor $g_{Z'}^4 / m_{Z'}^4$. We therefore consider a four-dimensional parameter space $(m_4,\, |U_{e 4}|^{2}, |U_{\mu 4}|^{2}, |U_{\tau 4}|^{2})$ and \cref{tab:BPs} shows some representative benchmark points used in this work.

\begin{table}
\def\arraystretch{1.5}
\begin{tabularx}{0.8\textwidth}{Y|YZYY}
  \toprule[1.2pt]
  & $m_{4}$ [GeV] & $\left|U_{e4}\right|^2$   & $\left|U_{\mu4}\right|^2$ & $\left|U_{\tau4}\right|^2$ \\ 
  \midrule[1pt]
  BP1a  & $2\times 10^{-3}$   &       0     & $9\times 10^{-3}$ &     0        \\
  BP1d & $2\times 10^{-3}$   &       0     & $9\times 10^{-3}$ & $9\times 10^{-3}$  \\
  \midrule[1pt]
  BP2a  & $9\times 10^{-3}$   &       0     & $9\times 10^{-3}$ &     0        \\
  BP2b & $9\times 10^{-3}$   &       0     & $9\times 10^{-3}$ & $9\times 10^{-4}$  \\
  BP2c & $9\times 10^{-3}$   &       0     & $9\times 10^{-3}$ & $4\times 10^{-3}$  \\
  BP2d & $9\times 10^{-3}$   &       0     & $9\times 10^{-3}$ & $9\times 10^{-3}$  \\
  \midrule[1pt]
  BP3a  & $20\times 10^{-3}$  &       0     & $9\times 10^{-3}$ &     0        \\
  \midrule[1pt]
  BP4a  & $40\times 10^{-3}$  &       0     & $9\times 10^{-3}$ &     0        \\ 
  \bottomrule[1.2pt]
  BP5a  & $60\times 10^{-3}$  &       0     & $9\times 10^{-3}$ &     0        \\ \bottomrule[1.2pt]
\end{tabularx}
\caption{Benchmark points used in this work.}
  \label{tab:BPs}
\end{table}

\section{Spallation source experiments}
\label{sec:spallation}

\begin{figure}
    \centering
    \includegraphics[scale = 0.6]{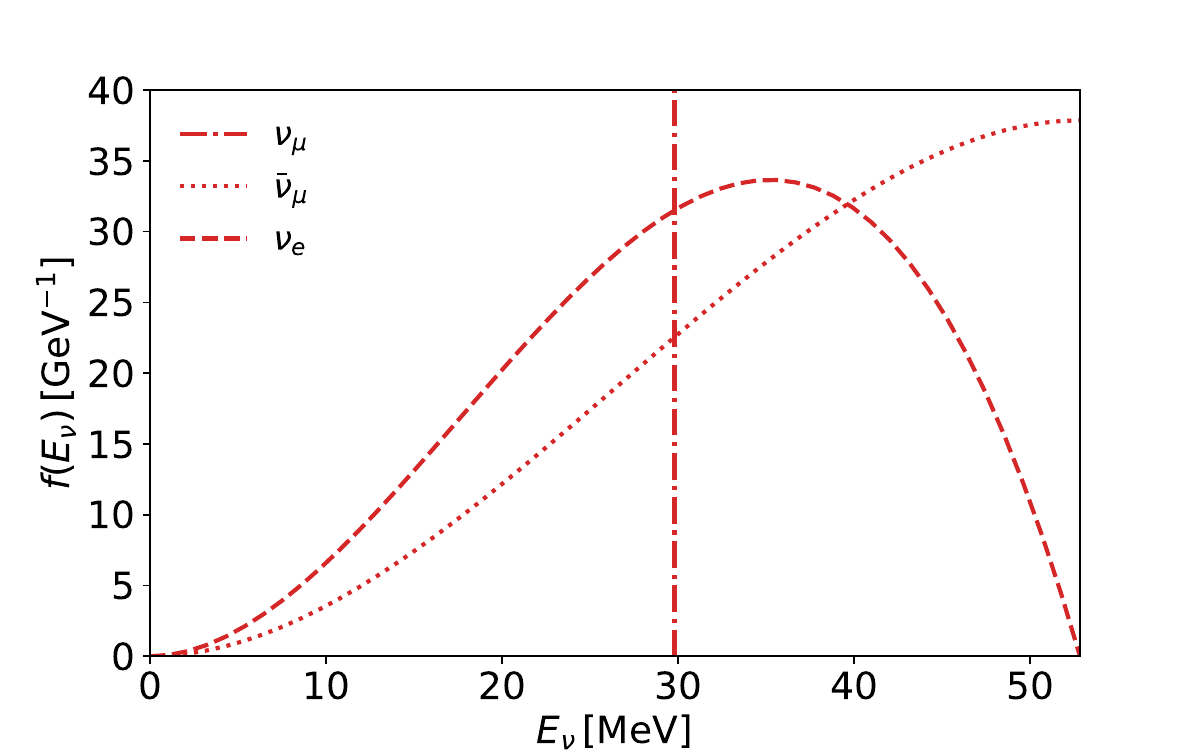}
    \caption{Normalised (without the $\xi$ factor) neutrino fluxes at spallation source facilities.}
    \label{fig:SS_flux}
\end{figure}

Neutrino experiments at spallation sources have become an extremely useful tool to explore new neutrino physics associated with neutrino-nucleus scattering. The neutrino flux arriving on-target has three components, shown in \cref{fig:SS_flux}. The prompt decay of the initially produced pions, $\pi^{+} \rightarrow \mu^{+} \nu_\mu$, induces a monochromatic beam of muon neutrinos with energy $E_{\nu_\mu} = (m_\pi^2 - m_\mu^2)/{2 m_\pi} \simeq \SI{30}{\mega\electronvolt}$. The delayed decay $\mu^+ \rightarrow e^+ \nu_e \bar{\nu}_\mu$ gives rise to a flux of muon antineutrinos and electron neutrinos with continuous energy distributions. The corresponding fluxes are given by (see, e.g., Ref.~\cite{Coloma:2017egw})
\begin{equation}
\begin{aligned}
    \diff{\phi_{\nu_{\mu}}}{E_\nu}&=
    \xi\delta\left(E_{\nu}-\frac{m_{\pi}^{2}-m_{\mu}^{2}}{2 m_{\pi}}\right)\,, \\
    \diff{\phi_{\bar{\nu}_{\mu}}}{E_\nu}&=
    \xi\frac{64}{m_{\mu}}\left[\left(\frac{E_{\nu}}{m_{\mu}}\right)^{2}\left(\frac{3}{4}-\frac{E_{\nu}}{m_{\mu}}\right)\right]\,, \\
    \diff{\phi_{\nu_{e}}}{E_\nu}&=
    \xi\frac{192}{m_{\mu}}\left[\left(\frac{E_{\nu}}{m_{\mu}}\right)^{2}\left(\frac{1}{2}-\frac{E_{\nu}}{m_{\mu}}\right)\right]\,,
    \end{aligned}
    \label{eq:coherent_flux}
\end{equation}
where, from kinematics, $E_\nu \in \left[0, m_\mu / 2\right]$ for the continuous spectra of $\bar{\nu}_{\mu}$ and ${\nu}_{e}$. The constant ${\xi \equiv r R_{\textup{POT}} / (4 \pi L^2)}$ accounts for the luminosity of the experiment. Here, $r$ is the number of neutrinos of any given flavour produced per proton collision, $R_{\textup{POT}}$ is the number of protons on target per unit time, and $L$ is the total length of the experimental baseline. Given the promising sensitivity of the configurations planned to run at the European Spallation Source, in this article we will consider it as a paradigmatic example of a realistic future experiment. Two different setups can be considered \cite{Baxter:2019mcx}: a small (10~kg) but extremely sensitive  detector with an energy threshold of $E_{\rm th}=0.1$~keV (which we refer to as ESS10), and a large detector (1~ton) but with a higher energy threshold of $E_{\rm th}=20$~keV (which we refer to as ESS). For both configurations, the baseline is $L=20$~m, $R_{\textup{POT}}=2.8\times10^{23}$~yr$^{-1}$, and $r=0.3$.  Despite the great advantage of its extremely low threshold, the small target size of ESS10 makes it insufficient to explore new regions of the parameter space of sterile neutrino models, and, for this reason, we will concentrate on ESS assuming 1~yr of operation. In our analysis, we consider a bin energy resolution of $5$~keV. For the quenching factor, we have extrapolated that of COHERENT-LAr \cite{COHERENT:2020iec}, $Q_F = 0.246+7.8\times10^{-4}E_R$, whereby $E[\si{\kilo\electronvolt}\ee{}]=Q_F\, E_R$. Following the treatment in Ref.~\cite{Amaral:2021rzw}, we approximate the efficiency as $\epsilon(E_R)=0.5\left(1+\tanh\left(E_R-E_{\rm th}\right)\right)/E_{\rm width}$, where we take $E_{\rm width}=1$~keV for ESS.

To compute the differential rate of nuclear recoil events, we integrate each neutrino flux, $\alpha' \in \left\{e,\mu,\bar\mu\right\}$, taking into account both SM \cevns and new physics up-scattering processes, from  \cref{eq:sigma_cevns} and \cref{eq:sigma_a4}, respectively. The differential scattering rate is given by
\begin{equation}
    \diff{R_{\alpha'}}{E_R} = \frac{1}{m_A} 
    \left( 
    \int_{E_{\nu}^{{\rm min, CE}\nu{\rm NS}}}
    ^{E_{\nu}^{\rm max}} 
    \diff{\phi_{\nu_{\alpha'}}}{E_\nu}
    \diff{\sigma_{{\rm CE}\nu{\rm NS}}}{E_R}\,\dl E_\nu\,
    +
    \int_{E_{\nu}^{\rm min, \alpha' 4}}^{E_{\nu}^{\rm max}}     \diff{\phi_{\nu_{\alpha'}}}{E_\nu} 
    \diff{\sigma_{\alpha' 4}}{E_R}\,\dl E_\nu\,
    \right)
    , 
    \label{eq:coherent_spectrum}
\end{equation}
where $1 / m_A$ is the total number of targets per unit mass in a given experiment, $\mathrm{d}{\sigma_{\bar\mu 4}}/\mathrm{d}{E_R}=\mathrm{d}\sigma_{\mu 4}/{\mathrm{d}E_R}$, and $E_{\nu}^{\rm max} = m_\mu /2$ is the maximum allowed neutrino energy. The minimum neutrino energy required to produce a recoil of energy $E_R$ differs for the elastic and inelastic processes. For  usual SM \cevns, it is given by
\begin{equation}
    E_{\nu}^{{\rm min,\, CE}\nu{\rm NS}} = \frac{1}{2}\left(E_{R}+\sqrt{E_{R}^{2}+2 m_A E_{R}}\right) \simeq \sqrt{\frac{m_A E_R}{2}}\,.
    \label{eq:nu_min_sm}
\end{equation}
However, for the inelastic up-scattering process, the minimum energy must be high enough to produce the massive sterile neutrino, leading to 
\begin{equation}
    E_{\nu}^{\rm min,\, \alpha' 4} = \left(1 + \frac{m_4^2}{2 m_A E_R}\right)E_{\nu}^{\min,\,\mathrm{CE\nu NS}}\,.
    \label{eq:nu_min_ndp}
\end{equation}

Finally, the total number of nuclear recoils in each energy bin is computed by integrating the differential rate over the experimental range of recoil energies (given by the specific experimental setup) weighted by the corresponding energy-dependent efficiency function, $\epsilon(E_R)$, 
\begin{equation}
    N_{\rm SS} = \varepsilon\sum_{\alpha'} \int_{E_R^{\rm min}}^{E_R^{\rm max}} \diff{R_{\alpha'}}{E_R}\epsilon(E_R)\, \dl E_R\ .
    \label{eq:coherent_n}
\end{equation}
where $\varepsilon$ is the experiment exposure: the product of its total mass and its live time. For the ESS configuration that we are considering, $\varepsilon=1\,\mathrm{ton}\,\mathrm{yr}$.

\begin{figure}
    \centering
    \includegraphics[scale = 0.6]{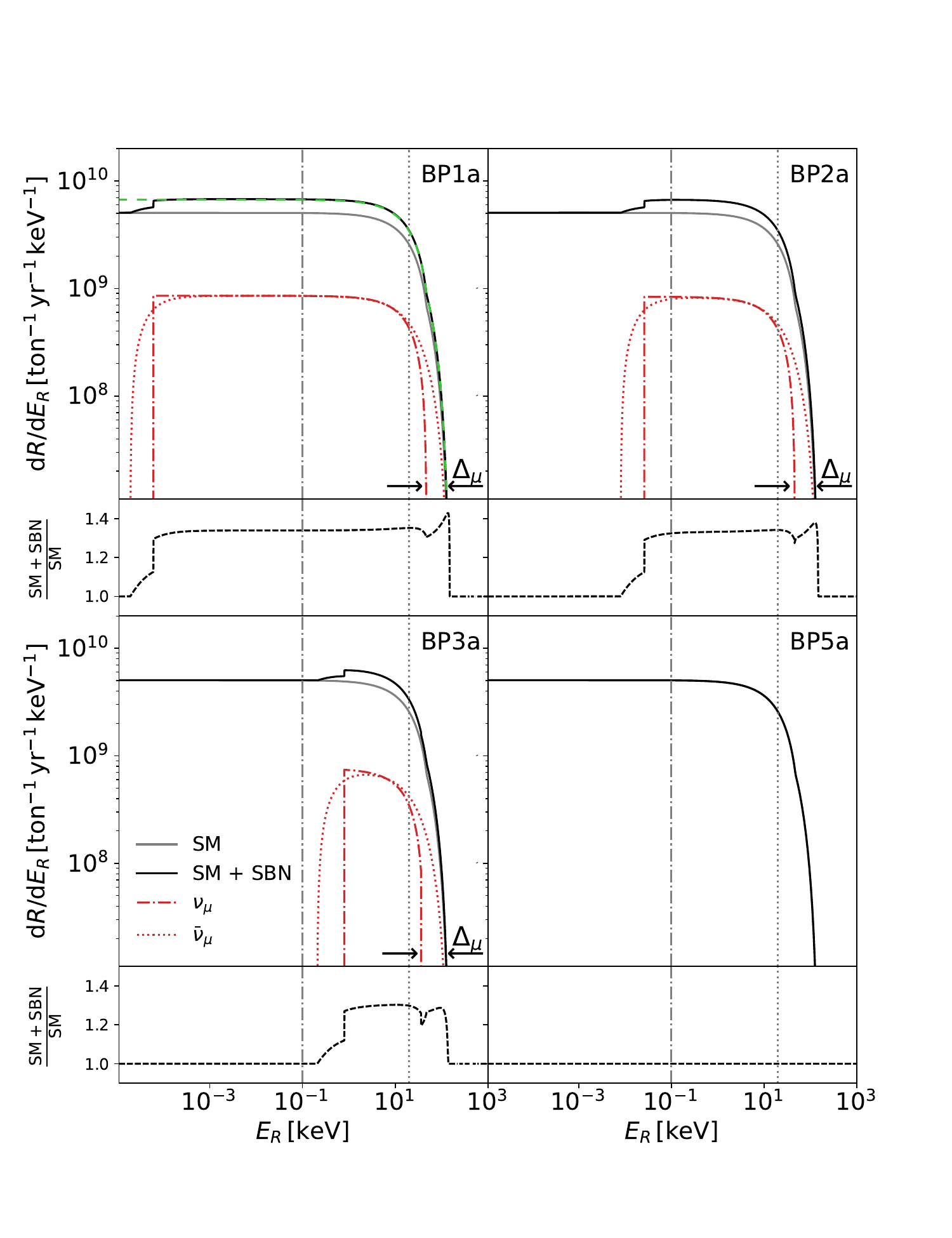}
    \caption{Predicted energy spectra at ESS for some representative benchmark points in \cref{tab:BPs}, featuring different values for $m_4$ and mixings with the active neutrinos. The vertical grey dotted (dashed-dotted) line shows the projected ESS (ESS10) threshold. The quantity $\Delta_\mu$ is defined as the energy difference between the endpoint of the SM spectrum and the contribution from the monochromatic $\nu_{\mu}$ flux. The dashed green line in the upper-left panel shows the expected neutrino NSI spectrum with $\varepsilon_{\mu\mu}^u = 0.4$.}
    \label{fig:SSspectra}
\end{figure}

\cref{fig:SSspectra} shows the differential spectrum for each contribution in \cref{eq:coherent_spectrum} and for four representative benchmark points (BP1a, BP2a, BP3a, and BP5a with parameters specified in \cref{tab:BPs}), where the sterile neutrino mass is varied for the same choice of couplings. The inelastic contribution only switches on above a certain recoil energy, leading to a characteristic bump with energies in the range
\begin{align}
    E_R^{\textup{bump}} \in& \left[  
    \frac{1}{2 m_A}\left(2 (E_{\nu}^{\rm max})^2 - m_4^2 - 2 E_{\nu}^{\rm max} \sqrt{ (E_{\nu}^{\rm max})^2 - m_4^2}\right)\right.\,,\\
   & \left.\frac{1}{2 m_A}\left(2 (E_{\nu}^{\rm max})^2 - m_4^2 + 2 E_{\nu}^{\rm max} \sqrt{ (E_{\nu}^{\rm max})^2 - m_4^2}\right)\right]\,,   
    \label{eq:E_bump}
\end{align}
where we have made the approximation $E_\nu/m_A\ll 1$. In the event of a future observation, this `bump' could be used to determine the mass of the sterile neutrino, thus helping to discriminate this model from other potential new physics contributions in the neutrino sector. In practice, this could confirm the existence of a sterile neutrino (with mass different from zero). Notice that the lower end of the energy bump takes place at very small values of the recoil energy, well below the reach of current and future detectors. For this reason, the sterile neutrino mass reconstruction mostly relies on determining the upper end of the bump, which is displaced from the end of the SM \cevns spectrum. The contribution from muon neutrinos is particularly interesting for this purpose. As their flux is monochromatic, the energy bump in their spectrum is more easily distinguishable from the SM prediction. The difference of the endpoint in the SM \cevns spectrum and the inelastic contribution from $\nu_\mu$ is denoted $\Delta_\mu$ in \cref{fig:SSspectra} for each benchmark point.

To observe this feature, the experimental threshold must be low enough and the energy resolution of the detector must at least be comparable to $\Delta_\mu$. Since $\Delta_\mu$ increases with $m_4$ (which we can see in \cref{fig:SSspectra} or infer from \cref{eq:E_bump}), heavier sterile neutrino masses are easier to reconstruct. Since the energy thresholds of current and planned experiments at spallation sources are of the order of $\sim 10$~keV, a measurement of the sterile neutrino mass is only possible above a certain value of $m_4$. In particular, given the planned characteristics of the ESS experiment, the signal of both BP1 and BP2 would be indistinguishable from that for $m_4=0$.
For reference, the vertical grey dotted (dashed-dotted) lines in \cref{fig:SSspectra} represent the expected energy threshold of both ESS and ESS10 respectively.

It should be emphasized that measuring the sterile neutrino mass---that is, confirming that $m_4=0$ is not within the $2\sigma$ best-fit region---is crucial to discriminate the signal due to the SBN model from that of a generic neutrino non-standard interaction (NSI), where no extra neutrinos are introduced \cite{Wolfenstein:1977ue,Guzzo:1991cp,Guzzo:1991hi,Gonzalez-Garcia:1998ryc,Bergmann:2000gp,Guzzo:2000kx,Guzzo:2001mi,GonzalezGarcia:2004wg}. Indeed, the spectrum from a particular choice of NSI can mimic the observed signal in the SBN model when the lower end of the energy bump is below the experimental threshold. We illustrate this in \cref{fig:SSspectra} for BP1a, where we have generated an NSI spectrum with a pure up-quark effective NSI parameter of $\varepsilon_{\mu\mu}^u = 0.4$. For the range of observable energies, we see that the SBN and NSI spectra almost completely overlap, making them indistinguishable from one another.

\begin{figure}[t]
    \centering
    \hspace*{-7ex}\includegraphics[width = 0.87\textwidth]{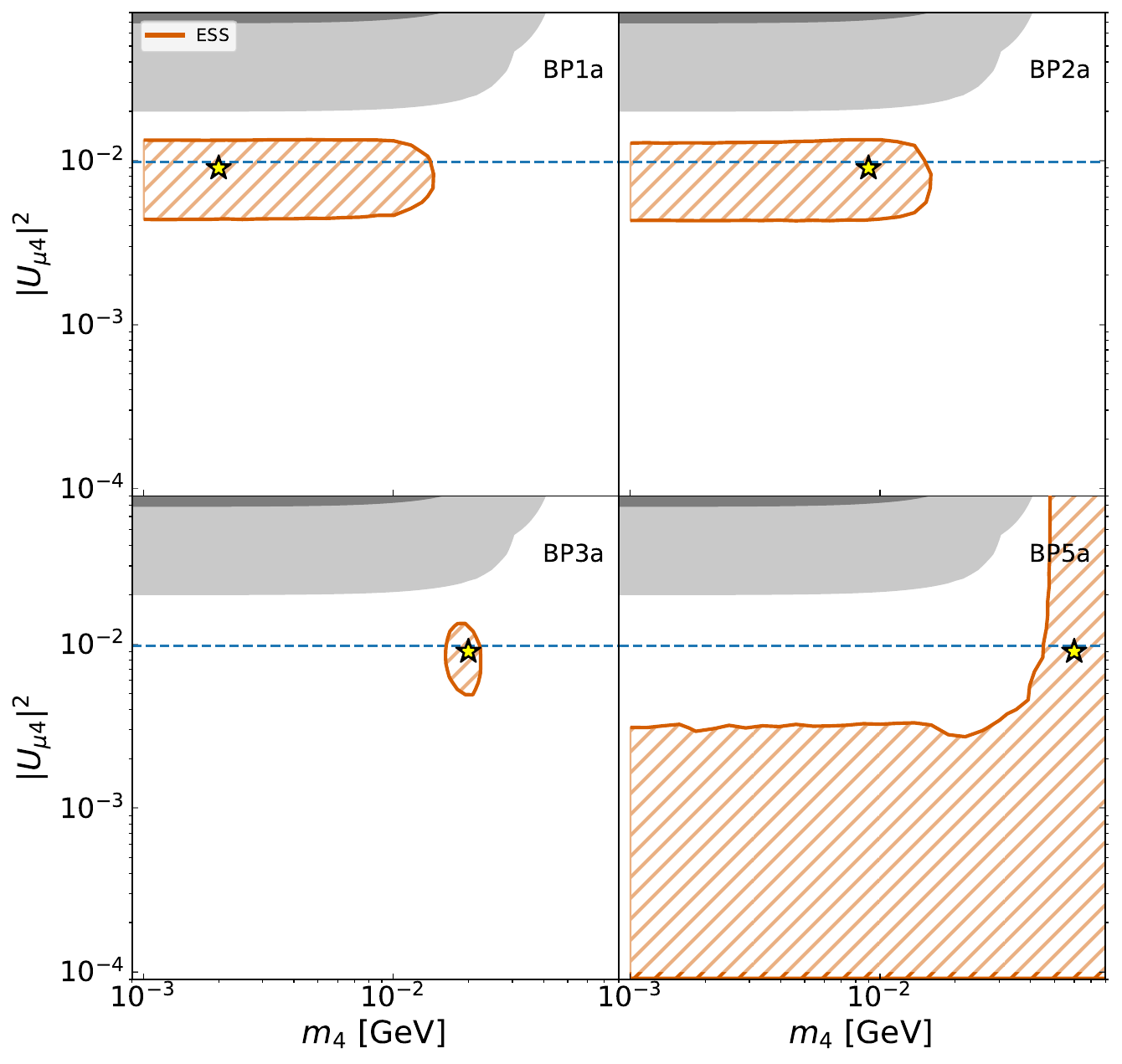}
    \caption{Profile-likelihood results for the BPs with $\left|U_{\mu4}\right|^{2}=9\times10^{-3}$, $\left|U_{\tau4}\right|^{2} = \left|U_{e4}\right|^{2} = 0$ and $m_{4}= 2$ MeV (\textit{upper left panel}), $9$ MeV (\textit{upper right panel}), $20$ MeV (\textit{lower left panel}), and $60$ MeV (\textit{lower right panel}) using SS data only. Hatched areas correspond to allowed regions ($\Delta \chi^{2} < 6.18$). For clarity, we depict as a yellow star the true values of the analysed benchmark point. The shaded black (grey) regions areas excluded by the current COHERENT data. The horizontal dashed blue line corresponds to the upper bound on the sterile neutrino mixing with the muon sector \citep{Arguelles_2023}. 
    } \label{fig:ESS_profile_likelihoods}
\end{figure}

To test the reconstruction of the sterile neutrino parameters, we have created Asimov data sets for each of these benchmark points and attempted to reconstruct their associated model parameters in the four-dimensional space $(m_4, \left|U_{e4}\right|^{2}, \left|U_{\mu 4}\right|^{2}, \left|U_{\tau 4}\right|^{2})$. In these Asimov  sets, our `observed' data are equal to the theoretically expected number of events for each given benchmark point. The ensuing limit from such an analysis should asymptotically approach the median limit arising from many Monte Carlo runs \cite{Cowan:2010js}. The statistical details of our analysis can be found in \cref{sec:app-stat}. We compute the expected number of nuclear recoil events from \cref{eq:coherent_n} using an extension of the \texttt{SNuDD} package \cite{SNuDD-code}. For each benchmark point, we carry out a profile-likelihood analysis using the nested sampling algorithm \texttt{multinest}  \cite{multinest,multinest2} via its Python implementation \cite{pymultinest}.

We show in \cref{fig:ESS_profile_likelihoods} the parameter reconstruction corresponding to BP1a, BP2a, BP3a, and BP5a, assuming the projected configuration of the ESS detector. The hatched areas correspond to the allowed regions ($\Delta \chi^{2} < 6.18$). As we can see, ESS would be able to observe the first three benchmark points and measure the coupling $\left|U_{\mu 4}\right|^{2}$. It would also be able to fully reconstruct the mass of the sterile neutrino in BP3a. Nevertheless, for BP1a and BP2a, only an upper bound on the sterile neutrino mass can be extracted (the end-point of the bump cannot be distinguished from the SM spectrum). Since the sterile neutrino mass for BP5a is above the energy of the neutrino flux in spallation source experiments, the up-scattering is kinematically forbidden and hence there will be no observation. For this benchmark point, we can only obtain an exclusion region.

As a new result, we have derived constraints on the SBN model using current COHERENT data from the two targets, LAr \cite{COHERENT:2020iec} and CsI \cite{COHERENT:2021xmm}. To do this, we have used the statistical treatment of \cref{sec:app-stat}. The bounds are represented in \cref{fig:ESS_profile_likelihoods} as light and dark grey areas in the corresponding plots for the LAr and CsI targets, respectively. As we can see, the excluded areas lie above the upper bound on the sterile neutrino mixing with the muon sector from Ref.~\cite{Arguelles_2023} and therefore do not probe new areas of the parameter space.

\begin{figure}[t]
    \centering
    \includegraphics[width=\textwidth]{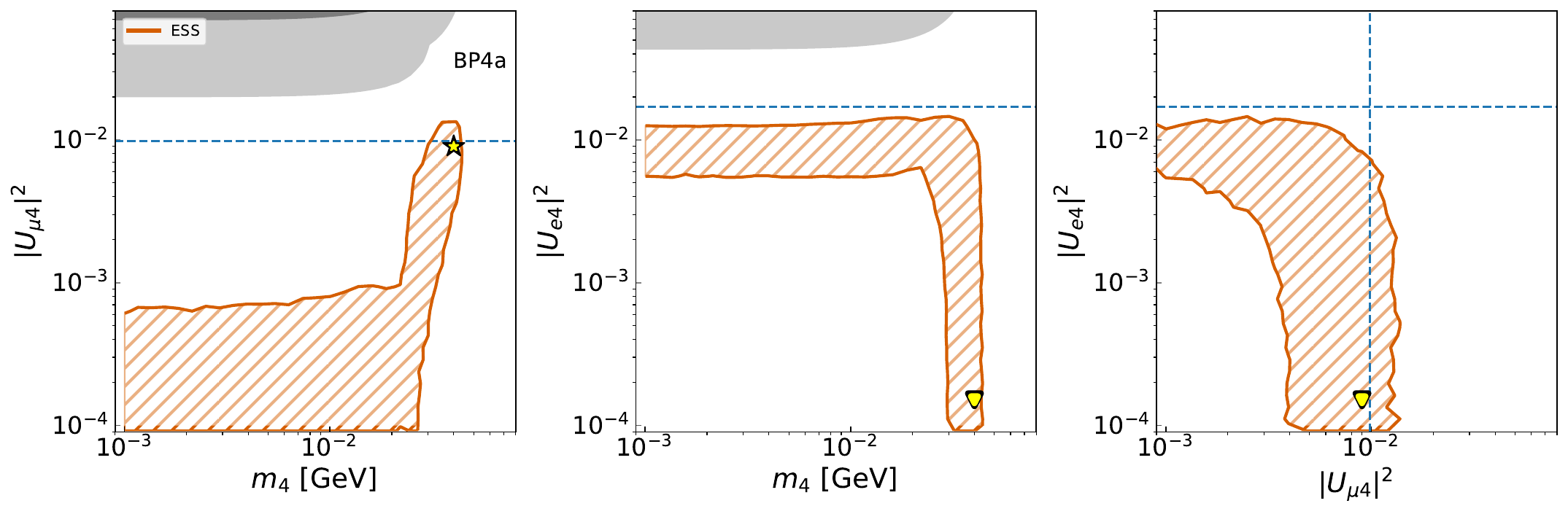}
    \caption{Profile-likelihood results for BP4a ($m_{4}=40$ MeV, $\left|U_{\mu4}\right|^{2} = 9\times10^{-3}$, $\left|U_{\tau4}\right|^{2} = \left|U_{e4}\right|^{2} = 0$) using SS data only. 
    For clarity, we depict as a yellow star the true values of the analysed benchmark point. The shaded black (grey) regions are excluded by current COHERENT data. The yellow arrows depict that the benchmark point is outside of the plotting area.
    } \label{fig:ESS_profile_likelihoods_BP4}
\end{figure}

It is interesting to note that for sterile neutrino masses above $m_4\gtrsim30$~MeV, the monochromatic $\nu_{\mu}$ flux is not energetic enough to produce the sterile neutrino and only the $\bar{\nu}_{\mu}$ and $\nu_{e}$ fluxes contribute in \cref{eq:coherent_spectrum}. When this occurs, the characteristic feature $\Delta_\mu$ is no longer present. This makes the mass reconstruction more difficult and leads to a degeneracy between the mixings with muon neutrinos, $U_{\mu4}$, and electron neutrinos, $U_{e4}$. This effect is more pronounced for $m_{4} \simeq 40$ MeV, where the $\nu_{e}$ and $\bar{\nu}_{\mu}$ fluxes are comparable. To exemplify this, in \cref{fig:ESS_profile_likelihoods_BP4} we analyse a benchmark point with $m_{4}=40$~MeV and $\left|U_{\mu 4}\right|^{2}=9 \times 10^{-3}$ (BP4a in \cref{tab:BPs}), which we attempt to reconstruct through a profile-likelihood analysis. The degeneracy on the reconstruction of the mixings (evidenced on the right panel) induces a similar degeneracy on the sterile neutrino mass (see left and middle panels of \cref{fig:ESS_profile_likelihoods_BP4}), making measuring $m_{4}$ impossible. This degeneracy is lifted for sterile neutrino masses $m_4 \gtrsim \SI{45}{\mega\electronvolt}$ (depending on the value of the mixings) when the contributions from the $\nu_{e}$ and $\nu_{\mu}$ fluxes differ (see \cref{fig:SS_flux}).

Our analysis so far shows that
\begin{itemize}
    \item Current limits on the SBN model using COHERENT data do not exclude new areas of the parameter space, but future experiments like ESS would allow us to explore regions below current experimental constraints.
    \item In the event of a positive observation, future SS experiments might be able to determine the sterile neutrino mass (distinguishing it from the massless case) for a range $m_4\sim 15-50$~MeV. For lighter masses, the observed signal is indistinguishable from that of a new massless neutrino.
    \item The sterile neutrino mixing with the electron and muon sectors can, in general, be disentangled based on the different shapes of the contribution from the $\nu_e$ and $\nu_\mu$ fluxes.
    \item There is, however, a region for sterile neutrino masses around $m_4 \sim \SI{40}{\mega\electronvolt}$ for which the reconstruction is highly degenerate and the sterile neutrino mass (and mixing with $\nu_e$ and $\nu_\mu$) cannot be measured.
    \item SS experiments are completely insensitive to the sterile neutrino mixing with the tau sector, as there is no $\nu_\tau$ flux.
\end{itemize}
In the following sections, we will study how (dark matter) direct detection experiments can provide complementary information that improves the reconstruction of the SBN model parameters, partially lifting some of these degeneracies and considerably improving the mass measurement.

\section{Direct Detection experiments}
\label{sec:direct}

While primarily employed in the search for dark matter, direct detection experiments are becoming so sensitive that they will start observing \cevns from solar neutrinos. Indeed, the sensitivities of xenon-based experiments of this and future generations---such as LZ \cite{LZ:2018qzl}, XENONnT \cite{XENON:2020kmp}, and DARWIN \cite{DARWIN:2016hyl}---are projected to hit the neutrino fog: a region of the parameter space where a dark matter signal and a neutrino event will be difficult to disentangle \cite{OHare:2021utq}. This motivates us to think of these experiments as neutrino observatories instead of as dark matter detectors, treating this `background' as a signal to help us learn more about the nature of both SM and BSM neutrino physics. In this section, we show how these experiments can use measurements of the solar neutrino scattering rate as a probe of the SNB model.

In the case of nuclear recoils, the calculation of the differential rate is similar to that of SS. The key differences are that we instead use the solar neutrino flux and that we must now account for the oscillation probabilities as neutrinos propagate to the Earth from the solar core. As we did in \cref{sec:spallation}, the SM and new inelastic contributions must be considered separately since the minimal neutrino energy to produce a nuclear recoil of a given energy differs. The differential scattering rate, after summing over the flavours $\alpha \in \{e,\, \mu,\, \tau\}$, is ultimately given by\footnote{It has recently been noted that one must be careful when calculating the solar neutrino scattering rate in the presence of new physics \cite{Coloma:2022umy}. If the new physics introduces flavour-changing neutral current processes, then a more general density matrix formalism must be employed. This was recently done in the context of DD experiments and general NSI in Ref.~\cite{Amaral:2023tbs}. In our case, flavour charge is conserved, so we can compute the rate in the usual manner as we have written.}
\begin{align}
    \diff{R}{E_R} = \frac{1}{m_A}
    &
    \left[
    \int_{E_{\nu}^{{\rm min,CE}\nu{\rm NS}}}
    ^{E_{\nu}^{\rm max}} 
    \diff{\phi_{\nu_e}}{E_\nu} 
    \diff{\sigma_{{\rm CE}\nu{\rm NS}}}{E_R}\,\dl E_\nu\, 
    +
    \sum_{\alpha} \int_{E_{\nu}^{\rm min, \alpha  4}}^{E_{\nu}^{\rm max}}     
    \diff{\phi_{\nu_e}}{E_\nu} \ P_{e\alpha}\ 
    \diff{\sigma_{\alpha 4}}{E_R}\,\dl E_\nu \right]\,
    , 
    \label{eq:dr_dd}
\end{align}
where ${\mathrm{d}\phi_{\nu_e}}/{\mathrm{d}E_\nu}$ is the total differential solar electron-neutrino flux and $P_{e\alpha}$ is the transition probability for an electron neutrino to oscillate to the flavour $\alpha$. Notice that since SM \cevns is flavour blind, the transition probabilities factor out and sum to one. For the new physics contribution, the cross section is instead flavour dependent, so the probabilities must be retained.

In this work, we consider a multi-ton xenon experiment with an exposure of $\varepsilon=200\,\mathrm{ton}\,\mathrm{yr}$, a recoil energy threshold of $E_{\mathrm{th}}=\SI{1}{\kilo\electronvolt}$, and an energy bin resolution of $\SI{1}{\kilo\electronvolt}$. This type of experiment has been shown to be a powerful probe of new physics in the neutrino sector \cite{Amaral:2023tbs,Amaral:2021rzw,Amaral:2020tga}. When calculating the total number of expected events, we incorporate experimental effects, folding into \cref{eq:dr_dd} the energy-dependent efficiency and resolution functions. We do this using
\begin{equation}
    \label{eq:n_dd}
    N_{\mathrm{DD}} = \varepsilon\int_{0}^{E_\mathrm{max}} \Bigg(\int \diff{R}{E'}\epsilon(E') \frac{1}{\sigma(E')\sqrt{2\pi}} \exp\left[-\frac{\left(E_R - E'\right)^2}{2\sigma^2(E')}\right]\,\dif E'\Bigg) \dif E_R\,,
\end{equation}
where the convolution with the Gaussian resolution function is taken with respect to the theoretically expected recoil energy, $E'$, which is converted to the observed recoil energy, $E_R$. The integral is taken from $E_R = 0$, with the threshold of the experiment implicitly incorporated through the efficiency function, $\epsilon$. Note that it is crucial to incorporate this convolution with the resolution function, as this smears lower energy $^8\mathrm{B}$ events beyond where \cevns would be kinematically forbidden. As experimental thresholds are typically placed near where this forbidden region occurs, which is useful for dark matter searches, this smearing allows us to see some events as opposed to almost no events.

To implement \cref{eq:dr_dd,eq:n_dd}, we once again make use of the \texttt{SNuDD} package. This package uses the B16-GS98 standard solar model neutrino flux predictions \cite{Vinyoles:2016djt} and the \texttt{NuFIT 5.2} oscillation parameter results to compute the electron neutrino survival and transition probabilities \cite{Esteban:2020cvm}. For more information on the package, please see Ref.~\cite{Amaral:2023tbs} for the theory and Ref.~\cite{SNuDD-code} for the code base.

With the existence of the new flavour state $\ket{\nu_b}$, it is possible that the electron neutrinos produced in the Sun can oscillate into baryonic neutrinos. These neutrinos could then \textit{elastically} scatter off target nuclei via the new vector mediator, leading to an observable signal in DD experiments that could, in principle, dominate over that of our considered inelastic process \cite{Pospelov:2011ha,Pospelov:2012gm}. However, for sterile neutrinos in the mass range we have considered ($m_4 \sim \SI{1}{\mega\electronvolt}\text{--}\SI{100}{\mega\electronvolt}$) deviations from the unitarity of the PMNS matrix are highly constrained by flavour and electroweak precision data, as well as direct searches for such heavy neutrino states \cite{Abdullahi:2022jlv}. Consequently, we take the liberty of ignoring transitions to the baryonic neutrino state, neglecting the elastic scattering process and using the SM prediction for the survival and transition probabilities.

\begin{figure}
    \centering
    \includegraphics[scale = 0.6]{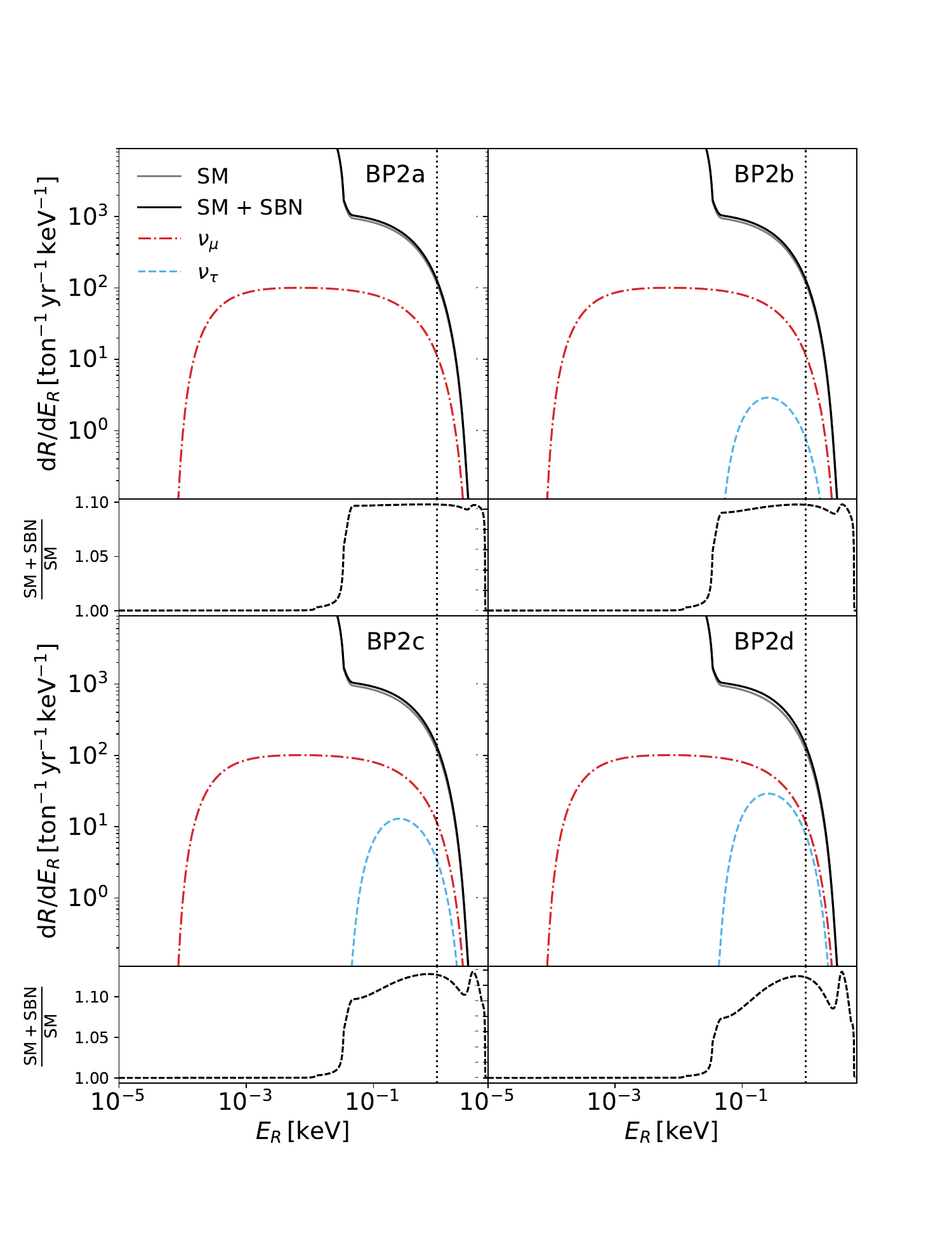}
    \caption{Predicted DD spectra for four benchmark points with $m_{4}=9$ MeV, $\left|U_{\mu4}\right|^{2} = 9\times 10^{-3}$, $\left|U_{e4}\right|^{2}=0$, and $\left|U_{\tau 4}\right|^{2} = 0~(\textit{upper\ left\ panel}) ,\ 9\times10^{-4}~(\textit{upper\ right\ panel}) ,\ 4\times10^{-3}~(\textit{lower\ left\ panel}) ,\ 9\times10^{-3}~(\textit{lower\ rigth\ panel})$.
    The SM spectrum is shown in grey, while the SBN contribution is shown in black. For completeness, we also show as red dashed-dotted (blue dashed) lines the SBN contributions arising from the $\nu_{\mu}$ ($\nu_{\tau}$) flux.}
    \label{fig:DDspectra}
\end{figure}

\cref{fig:DDspectra} shows the resulting differential spectrum for some representative benchmark points from \cref{tab:BPs}. As in the case of SS experiments, the new physics contribution from the inelastic process shows a characteristic bump. There is, however, an important difference. Since the solar neutrino fluxes are not monochromatic, this feature is not as abrupt as the $\nu_{\mu}$ contribution in SS experiments. Consequently, the reconstruction of the sterile neutrino mass from a hypothetical future signal in DD experiments is significantly more challenging. Notice that the lower end of the energy bump is generally well below the experimental threshold (and is therefore not observable). Thus, it is difficult to determine a lower bound on the mass of the sterile neutrino using DD alone. Given the shape of the solar neutrino flux \cite{Bahcall:2004pz}, for sterile neutrino masses above $\sim \SI{2}{\mega\electronvolt}$, only the $^8$B and $hep$ neutrino fluxes contribute to the inelastic process.
Despite this, DD experiments have the great advantage that they are sensitive to all three flavours of active neutrinos, thereby conveniently complementing the information from spallation sources, which lack a tau neutrino flux.

As we did for SS experiments, we can compare the expected number of events for a given set of model parameters with the simulated data of each benchmark point detailed in \cref{tab:BPs}. Since the expected number of events is significantly lower than in SS experiments, we model the likelihood as a product of Poissonian likelihoods for each energy bin. In addition, we introduce a nuisance parameter to account for the systematic uncertainty on the $^8 \mathrm{B}$ flux. The full statistical description can be found in \cref{sec:app-stat}. To test how this uncertainty impacts our results, we consider two cases\footnote{These values are motivated by the current uncertainty obtained through global fits analysis \cite{Bergstrom:2016cbh} ($\sigma_{^8 \mathrm{B}} = 2\%$) and the uncertainty to which DUNE will measure $\mathrm{^{8}B}$ using a combination of  elastic scattering and charged-current interactions ($\sigma_{^8 \mathrm{B}} = 2.5\%$) \cite{Capozzi:2018dat}.}: one with the current experimental uncertainty of $\sigma_{^8 \mathrm{B}} = 4\%$ \citep{SNO:2011hx} and another one with an optimistic uncertainty of $\sigma_{^8 \mathrm{B}} = 1\%$.

In \cref{fig:DDSS_profile_likelihoods}, we show as blue hatched regions the parameters that would be allowed ($\Delta \chi ^2 < 6.18$) by a future observation in a multi-ton liquid xenon experiment with $\sigma_{^8 \mathrm{B}} = 1\%$. For comparison, we include as a blue dashed line the results obtained with $\sigma_{^8 \mathrm{B}} = 4\%$. Given the maximum energy of the $^8$B solar neutrino flux, DD experiments will be insensitive to BP3a and BP5a. Hence, DD experiments can only probe sterile neutrinos with a low mass ($m_4\lesssim 20$~MeV) and a large mixing. Regarding the benchmark points of \cref{fig:DDSS_profile_likelihoods}, only BP1a is observable---while we do observe events for BP2a, the statistics are not high enough for a reconstruction. For BP2a, BP3a, and BP5a we only obtain an upper bound on the neutrino mixing. For BP5a, adding DD data leads to a more constraining upper bound for small sterile neutrino masses. It should be emphasised that one cannot disentangle the individual contributions from each of the three neutrino flavours using only DD data, and therefore the reconstruction of the mixing parameters is completely degenerate (in the figure, this leads to $|U_{\mu 4}|^2$ being unbounded).

\section{The Complementarity of direct detection and spallation source Experiments} \label{sec:complementarity}

In this section, we forecast the sensitivity that will be achieved by combining the results of future DD and SS experiments. In particular, we analyse how their complementarity can be used to break the degeneracies found in their individual analyses and better determine the parameters of the SBN model. Since the measurements performed by DD and SS experiments are independent of one another, we model the total likelihood as the product of the individual likelihoods described in \cref{sec:app-stat}. Using this combined likelihood, we repeat our previous analysis.

\begin{figure}[!t]
    \centering
    \includegraphics[width = 0.87\textwidth]{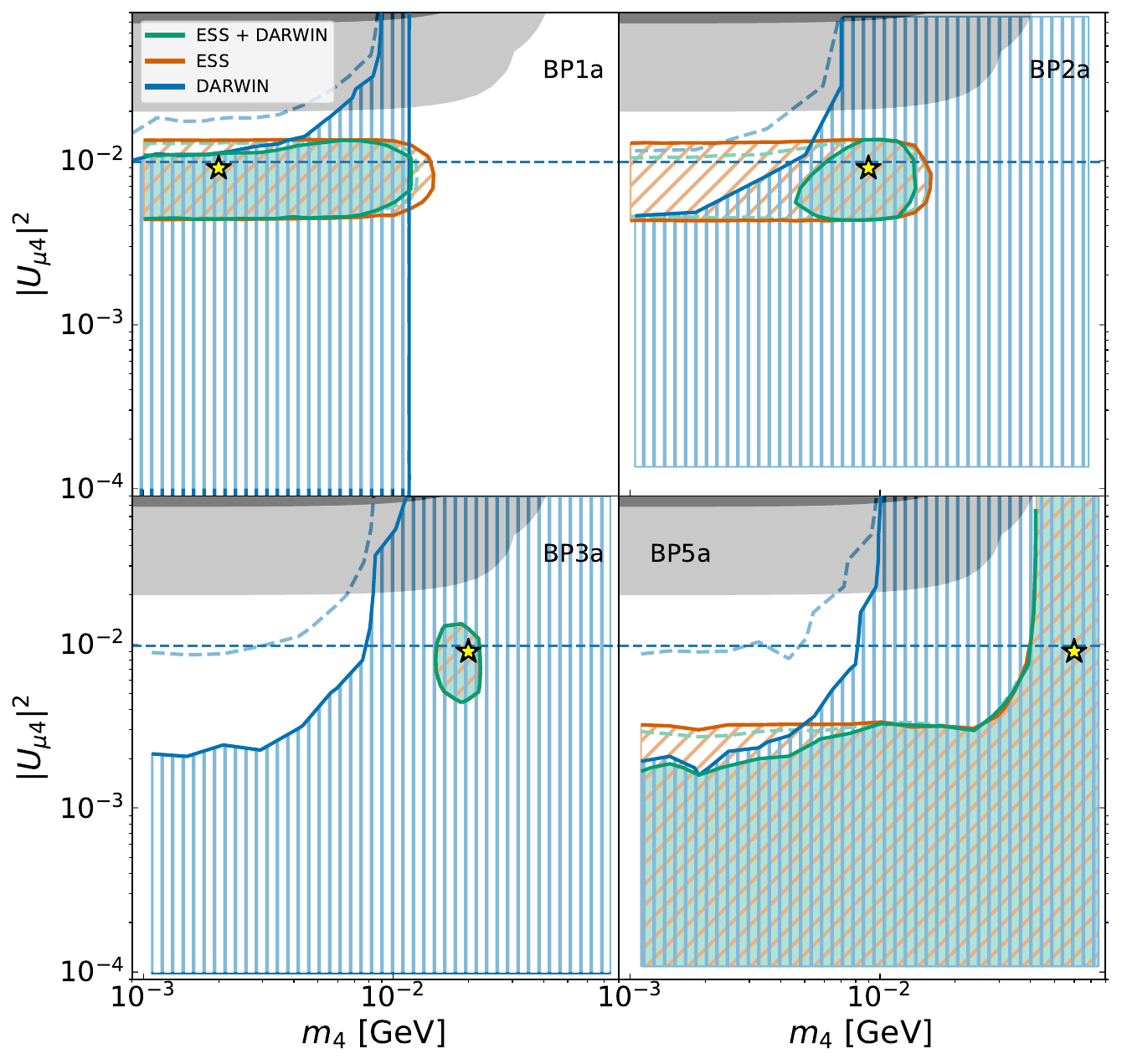}
    \caption{Profile-likelihood results for the BPs with $\left|U_{\mu4}\right|^{2}=9\times10^{-3}$, $\left|U_{\tau4}\right|^{2} = \left|U_{e4}\right|^{2} = 0$, and $m_{4}= 2$ MeV (\textit{upper left panel}), $9$ MeV (\textit{upper right panel}), $20$ MeV (\textit{lower left panel}), and $60$ MeV (\textit{lower right panel}). The orange (blue) hatched regions show the results using SS (DD) data only, while the green shaded regions show the results when using the combination of both types of experiments. For clarity, we depict as a yellow star the true values of the analysed benchmark points. The shaded black (grey) regions are excluded by current COHERENT data. Regarding the uncertainty in the $^8$B solar neutrino flux, the solid blue line corresponds to $\sigma_{^8 \mathrm{B}} = 1\%$, and the dashed blue line to $\sigma_{^8 \mathrm{B}} = 4\%$. }
    \label{fig:DDSS_profile_likelihoods}
\end{figure}

In \cref{fig:DDSS_profile_likelihoods}, we present the results for the same benchmark points as in \cref{fig:ESS_profile_likelihoods}, but now considering the information that DD experiments can contribute. The blue-shaded areas correspond to the best-fit regions when only DD data are considered, while green-shaded regions are those that employ the combination of DD and SS data. Only BP1a is observable by a future multi-ton xenon experiment. While the corresponding mass of BP1a cannot be determined using DD alone, the inclusion of DD data leads to a more stringent upper bound on $m_4$. For BP2a, BP3a, and BP5a, DD can only set upper bounds on the mixing parameters; however, this can still prove to be extremely useful. For example, when combined with SS results, this can help to exclude regions with small $m_4$. In the case of BP2a, for instance, DD  complements the results of SS and is crucial to better measure the sterile neutrino mass. For BP5a, DD data improves the exclusion for small values of $m_4$.

\begin{figure}
    \centering
    \includegraphics[width=\textwidth]{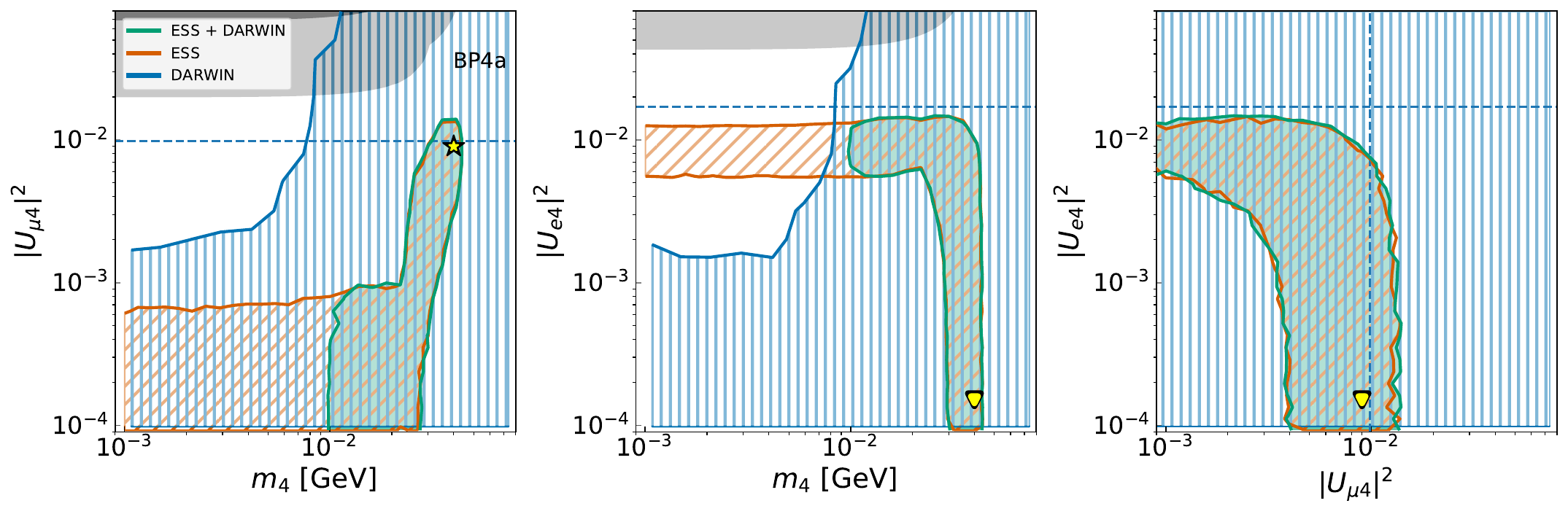}
    \caption{Profile-likelihood results for BP4a ($m_{4}=40$ MeV,  $\left|U_{\mu4}\right|^{2} = 9\times10^{-3}$, $\left|U_{\tau4}\right|^{2} = \left|U_{e4}\right|^{2} = 0$). The orange (blue) hatched regions show the results of using SS (DD) data only, while the green-shaded regions show the results when using the combination of both types of experiment. For clarity, we depict as a yellow star the true values of the analysed benchmark point. The shaded black (grey) regions areas are excluded by current COHERENT data. The yellow arrows depict that the benchmark point is outside of the plotting area.
    } \label{fig:FULL_profile_likelihoods_BP4}
\end{figure}

A particularly interesting case is that of BP4a. As explained in \cref{sec:spallation}, for $m_{4} \simeq \SI{40}{\mega\electronvolt}$, the parameter reconstruction using only data from SS experiments displays a degeneracy in the sterile neutrino mixings and mass (see \cref{fig:ESS_profile_likelihoods_BP4}). In \cref{fig:FULL_profile_likelihoods_BP4}, we show how this degeneracy is partially lifted when DD data is included. Although BP4a is not observable in a future xenon detector because of its large mass, the bounds from DD exclude the region of the parameter space with small $m_4$ and large $\left|U_{e4}\right|^{2}$, which in turn leads to a good measurement of the sterile neutrino mass.

\begin{figure}
    \centering
    \includegraphics[width=\textwidth]{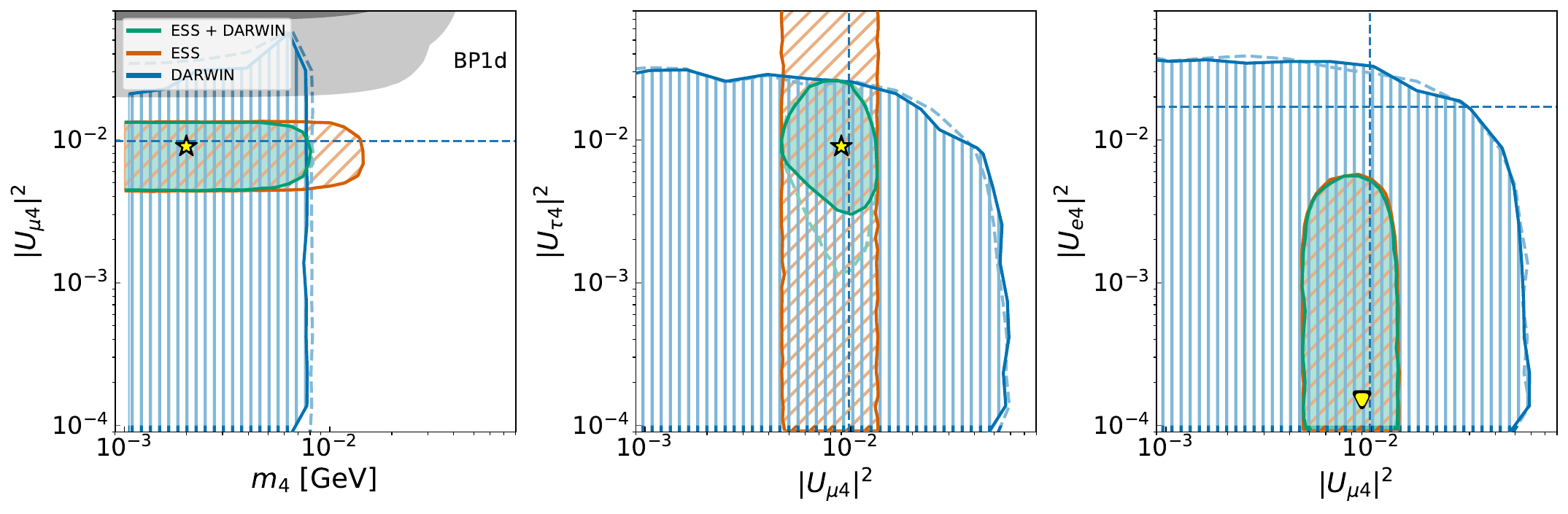}
    \caption{
    The same as in \cref{fig:FULL_profile_likelihoods_BP4} but for BP1d.}
    \label{fig:profile_likelihood_BP1b_c_d}
\end{figure}

Another great advantage of combining both types of experiments is that the solar neutrino flux includes a $\nu_\tau$ component due to neutrino oscillations. This provides an extra handle with which to measure the sterile neutrino mixing with tau neutrinos. In order to test this, \cref{fig:profile_likelihood_BP1b_c_d} shows an analysis of BP1d: a benchmark point with a non-negligible $U_{\tau4}$ mixing. Not only is this component measured with DD data, but also the combination with SS results leads to a better upper bound on the sterile neutrino mass and an improved reconstruction of $U_{\tau4}$.

\begin{figure}
    \includegraphics[width=\textwidth]{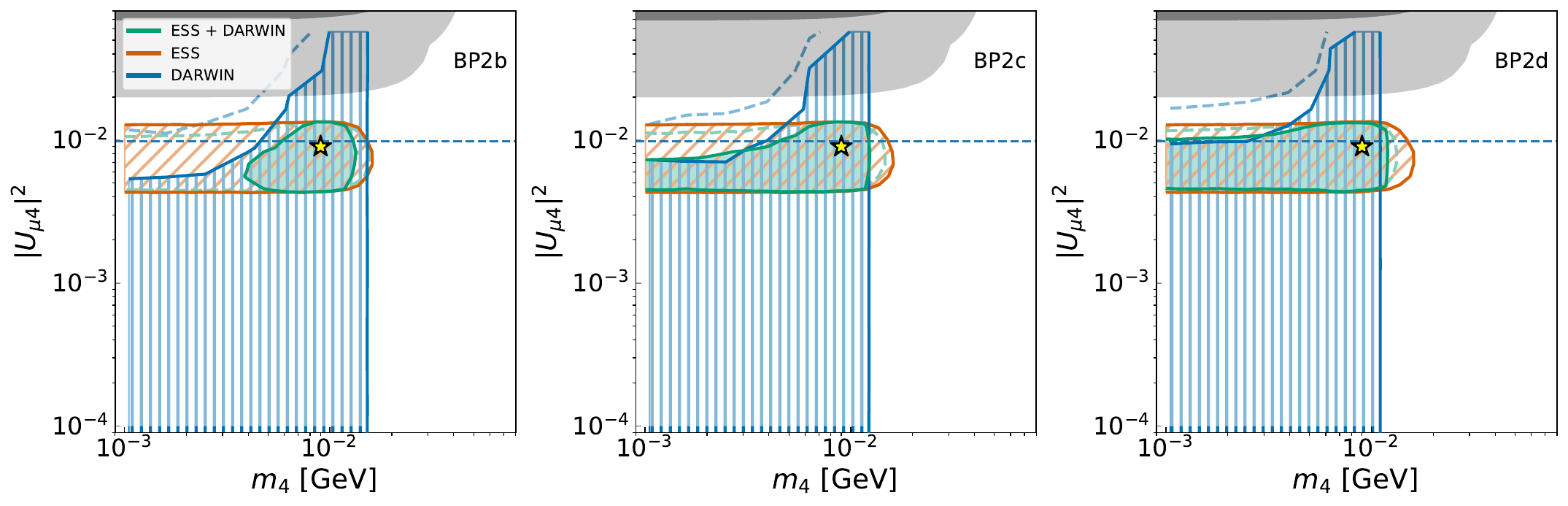}
    \caption{
   Profile-likelihood results for benchmark points with $m_{4}=2$ MeV, $\left|U_{\mu4}\right|^{2}=9 \times 10^{-3}$ and $\left|U_{\tau4}\right|^{2} = 9\times10^{-4} ~(\textit{left}) ;\ 4\times10^{-3}~(\textit{middle}) ;\ 9\times10^{-3}~(\textit{right})$.  The orange (blue) hatched regions show the results using SS (DD) data only, while the green shaded regions show the results when using the combination of both types of experiments. For clarity, we depict as a yellow star the true values of the analysed benchmark points. The shaded black (grey) regions are excluded by current COHERENT data.
   }
    \label{fig:BP2b}
\end{figure}

For completeness, \cref{fig:BP2b} shows a series of examples where both $U_{\mu 4}$ and $U_{\tau 4}$ are non-vanishing, corresponding to BP2b, BP2c, and BP2d in \cref{tab:BPs}. These benchmark points are observable in DD thanks to the $U_{\tau 4}$ component. When the best-fit regions are determined, the upper bound on $|U_{\mu 4}|^2$ from DD data is sensitive to the magnitude of the mixing with tau neutrinos: for small $|U_{\mu 4}|^2$ (e.g., BP2b), the bound on $|U_{\mu 4}|^2$ is less stringent than when $|U_{\mu 4}|^2$ increases (e.g., BP2d). This also makes the combination with SS results less trivial---in some cases, the excluded regions allow for a better reconstruction of the sterile neutrino mass (BP2b), whereas in other cases this is not possible (BP2c and BP2d).

\subsection{How well can we measure the sterile neutrino mass?}

As we have demonstrated, the combination of DD data with that from SS experiments can lead to a better measurement of the sterile neutrino mass. This can happen even in the cases where DD would not observe a new physics signal, simply from the effect that the DD exclusions have on the regions of the parameter space that are consistent with detection in SS experiments. Reconstructing $m_4$ (i.e., confirming that it is non-vanishing) is crucial to discriminate a sterile neutrino model from other kinds of BSM neutrino physics (such as NSI on the active neutrinos).

\begin{figure}
    \centering
    \includegraphics[width = 0.87\textwidth]{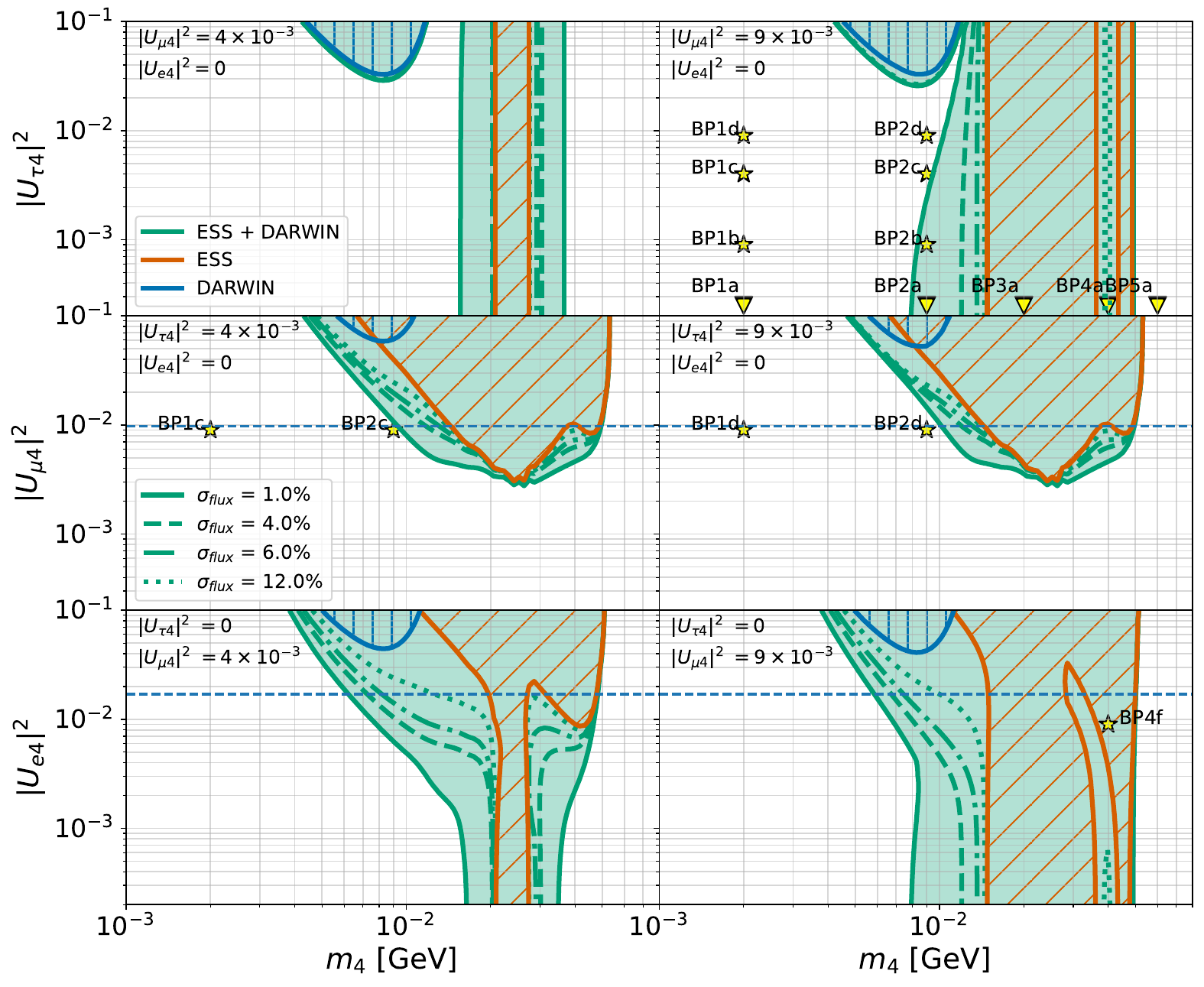}
    \caption{Regions of the parameter space in which we have a $2\sigma$ mass reconstruction for different solar flux uncertainties. 
    The orange (blue) hatched regions show the results using SS (DD) data only, while the green shaded regions show the results when using the combination of both types of experiments.
    \textit{Upper panel:} Results for benchmark points with fix $\left|U_{e4}\right|^{2} = 0$ and $\left|U_{\mu4}\right|^{2} = 4\times 10^{-3} (9\times 10^{-3})$ in left and right column respectively.
    \textit{Middle panel:} Results for benchmark points with fix $\left|U_{e4}\right|^{2} = 0$ and $\left|U_{\tau4}\right|^{2} = 4\times 10^{-3} (9\times 10^{-3})$ in left and right column respectively.
    \textit{Lower panel:} Results for benchmark points with fix $\left|U_{\tau4}\right|^{2} = 0$ and $\left|U_{\mu4}\right|^{2} = 4\times 10^{-3} (9\times 10^{-3})$ in left and right column respectively. The yellow arrows depict that the benchmark point is outside of the plotting area. 
    }
    \label{fig:Deltachi2FULL}
\end{figure}

In order to better quantify the relevance of the DD and SS complementary role in measuring $m_4$ and to provide a more general picture, we show in \cref{fig:Deltachi2FULL} various projections of the $(m_4,\,|U_{e4}|^{2},\,|U_{\mu 4}|^{2},\,|U_{\tau 4}|^{2})$ parameter space, indicating the areas where $m_4$ can be reconstructed (i.e., $m_4=0$ is not within the $95\%$ CL region). Using the same colour convention as in previous plots, the orange (blue) areas are those where $m_4$ can be reconstructed solely from SS (DD) data, and green regions correspond to their combination. From top to bottom, the first row corresponds to the  $(m_4,\,|U_{\tau 4}|^{2})$ plane with $|U_{e4}|^{2} = 0$ and $|U_{\mu4}|^{2} = 4 \times 10^{-3} ~  (9 \times 10^{-3})$ left (right) column. The second row shows the $(m_4,\,|U_{\mu 4}|^{2})$ plane with $|U_{e4}|^{2} = 0$ and $|U_{\tau4}|^{2} = 4 \times 10^{-3} ~ (9 \times 10^{-3})$ left (right) column. In the third row, we represent the $(m_4,\,|U_{e 4}|^{2})$ plane for $|U_{\tau4}|^{2} = 0$ and $|U_{\mu4}|^{2} = 4 \times 10^{-3} ~ (9 \times 10^{-3})$ left (right) column. The different benchmark points of \cref{tab:BPs} are indicted with yellow stars.

In all of these figures, we observe a clear synergy between DD and SS experiments. This is evinced by the green areas extending beyond the union of the blue and orange ones. In particular, the addition of DD data allows us to measure smaller values of $m_4$. The gap in the orange area of the top right and lower right panels appears for $m_4\simeq 40$~MeV and corresponds to the regions where the degeneracy between $|U_{e4}|^{2}$ and $|U_{\mu4}|^{2}$ makes the mass reconstruction impossible for SS experiments alone (see \cref{fig:ESS_profile_likelihoods_BP4} for BP4a). The addition of DD information is crucial to break this degeneracy and, hence, allow  for a mass reconstruction in this region (as in \cref{fig:FULL_profile_likelihoods_BP4}).

As already mentioned, the performance of DD experiments is extremely sensitive to the uncertainty in the solar neutrino fluxes. For completeness, in \cref{fig:Deltachi2FULL} we show in dashed, dashed-dotted and dotted green lines the results obtained when combining both types of experiments and considering a $^8\mathrm{B}$ flux uncertainty of $4\%$, $6\%$ and $12\%$, respectively. As expected, we see how our results worsen when increasing this uncertainty.

\section{Conclusions}
\label{sec:conclusions}

In this work, we have analysed the complementarity of direct detection and spallation source experiments for the study of sterile neutrino physics. Specifically, we have focused on the sterile baryonic neutrino (SBN) model: an extension of the SM that incorporates a new gauge boson that couples to baryons and a sterile neutrino that mixes with the active ones and also couples to this mediator. Due to this mixing, the sterile neutrino can be produced through the up-scattering of an active neutrino with the nucleus of a target material. This inelastic process alters the expected nuclear recoil spectra for both DD and SS experiments, providing a characteristic signature that can allow for the measurement of the sterile neutrino mass and mixing parameters in the event of a future detection.

Using current data from the COHERENT collaboration on CsI and LAr, we have first derived new constraints on the SBN model, showing that they do not exclude new areas of the parameter space. Assuming a future SS experiment with the projected properties of a detector to be installed at the ESS, we have then assessed how well the sterile neutrino properties would be determined upon a positive observation. We have shown that  the new inelastic contribution to neutrino-nucleus scattering induces a bump in the nuclear recoil spectrum. This proves extremely useful to reconstruct the sterile neutrino mass, conclusively disentangling this model from a generic NSI contribution to the active neutrinos. We have demonstrated that using only SS data, values in the range $15-50$~MeV can be measured. However, in a narrow range of masses of the order of 40~MeV, there is a degeneracy in the measurement of the sterile neutrino mixing that substantially affects mass reconstruction.

Incorporating future DD data helps in two ways. These detectors have an excellent energy resolution and generally a lower energy threshold than SS experiments. Furthermore, DD experiments are sensitive to all three neutrino flavours, including tau neutrinos, present in the solar neutrino flux. Thus, they are extremely helpful in removing degenerate solutions in the neutrino mixing parameter space. Considering the case of a future multi-ton liquid xenon experiment, we have demonstrated that the combination of future DD and SS results is crucial to substantially increase the area of the parameter space where the sterile neutrino mass can be reconstructed (see \cref{fig:Deltachi2FULL}), allowing us to measure values as low as $\sim 8$~MeV.

These results strengthen the role of DD experiments as probes of the neutrino sector and their complementarity with dedicated neutrino detectors.

\section*{acknowledgements}

We would like to thank Pilar Coloma, Manuel Gonz\'alez-L\'opez, El\'\i as L\'opez Asamar, Patrick Foldenauer, Marina Cerme\~no, Andr\'es P\'erez and Karen Mac\'\i as for useful discussions and comments. DAG, DGC and MdlR acknowledge support from the Comunidad Autonoma de Madrid and Universidad Autonoma de Madrid under grant SI2/PBG/2020-00005, and by the Spanish Agencia Estatal de Investigaci\'on through the grants PID2021-125331NB-I00 and CEX2020-001007-S, funded by MCIN/AEI/10.13039/501100011033. DGC also acknowledges support from the Spanish Ministerio de Ciencia e Innovaci\'on under grant CNS2022-135702.
DA is supported by the National Science Foundation under award 2209444.

\appendix

\section{Statistical Treatment}
\label{sec:app-stat}

In all of our analyses, we consider the profiled log-likelihood-ratio test statistic, defined as
\begin{equation}
  q(\vec{\theta;\, \vec{\zeta}_0}) \equiv -2 \ln \left[\frac{\mathcal{L}(\vec{\theta},\, \hat{\vec{\omega}},\, \hat{a};\,\vec{\zeta}_0)}{\mathcal{L}(\hat{\hat{\vec{\theta}}},\, \hat{\hat{\vec{\omega}}},\, \hat{\hat{a}};\,\vec{\zeta}_0)}\right]\,,
  \label{eq:llr}
\end{equation}
where $\mathcal{L}$ is the likelihood function describing our data given the model parameters. For later convenience, we have split our model parameters into three subsets, represented by $\vec{\theta}$, $\vec{\omega}$, and $\vec{\zeta}_0$. The parameters $\vec{\theta} \equiv (m_4,\, |U_{\alpha 4}|^{2})^{\mathrm{T}}$, for some given flavour index $\alpha \in \{e,\,\mu,\,\tau\}$, are the two parameters we are constraining at any given time. The parameters $\vec{\omega} \equiv (|U_{\beta 4}|^{2},\, |U_{\gamma 4}|^{2})^\mathrm{T}$, with $\alpha \neq \beta \neq \gamma$, are the two remaining mixings we profile over at a given BP. Finally, as explained in \cref{sec:sbn}, we fix the parameters related to the new vector mediator, denoted by $\vec{\zeta}_0 \equiv (g_{Z'},\,m_{Z'})^{\mathrm{T}}$. We also introduce a dimensionless pull parameter, $a$, as a nuisance parameter that is designed to capture systematic uncertainties in the theoretically expected count. We model this parameter as being Gaussian distributed with a mean of zero and an experiment-dependent standard deviation.
Hatted variables indicate quantities that maximise the likelihood at a given parameter space point (the null hypothesis likelihood), while double-hatted variables represent the quantities that maximise the unconstrained likelihood (that of the alternative hypothesis).

\subsection{Spallation Source Experiments}
\label{subsec:app_spallation}

Following Refs.~\citep{COHERENT:2017ipa,Miranda:2020tif,Amaral:2021rzw} for SS experiments, we perform a binned statistical analysis, modelling the likelihood of each bin $i$ as a Gaussian. In this case, \cref{eq:llr} reduces to the simpler $\Delta\chi^2$ statistic, with
\begin{equation}
  \chi^2(\vec{\theta},\, \vec{\omega},\,a;\,\vec{\zeta}_0) = \sum_{i = 1}^{N_\mathrm{bins}}\left(\frac{N^i_\mathrm{obs} - \left[1 + a\right]N_\mathrm{th}^i(\vec{\theta},\, \vec{\omega};\, \vec{\zeta}_0)}{\sigma_\mathrm{stat}^i}\right)^2 + \left(\frac{a}{\sigma_\mathrm{sys}}\right)^2\,.
  \label{eq:chi2-ss}
\end{equation}
Here, $N^i_\mathrm{obs}$ and $N_\mathrm{th}^i(\vec{\theta},\,\vec{\omega};\, \vec{\zeta}_0)$ are the numbers of observed and theoretically expected events in the $i^\mathrm{th}$ bin, respectively. The quantity $\sigma_\mathrm{stat}^i$ is the statistical uncertainty of the observed number of events, which we take to be
\begin{equation}
  \sigma_\mathrm{stat}^i \equiv \sqrt{N_\mathrm{obs}^i + N_\mathrm{bkg}^i}\,,
\end{equation}
where $N_\mathrm{bkg}^i$ is the expected number of background events in the $i^\mathrm{th}$ bin. When performing our analysis of COHERENT data, we use the backgrounds reported by collaboration \citep{COHERENT:2020iec,COHERENT:2021xmm}. However, when considering the future ESS experiment, we instead use the fact that the beam-related neutron (BRN) background represents an important background in this type of search, with CENNS-10 reporting that $10\%$ of its measured signal events arose due to this background source \citep{COHERENT:2019kwz}. Since we make no assumptions on how well future SS experiments will handle this background, we take $N_\mathrm{bkg}^i \equiv N_\mathrm{SM}^i / 10$, with $N_\mathrm{SM}^i$ the number of expected \cevns events in the $i^\mathrm{th}$ bin under the SM. For the pull parameter, $a$, we take its uncertainty to be $\sigma_\mathrm{sys} = 0.05$ \citep{Miranda:2020syh,Amaral:2021rzw}.

To construct the $\Delta \chi^2$ for our parameters of interest, we compute the profiled test statistic
\begin{equation}
    \Delta \chi^2(\vec{\theta};\,\vec{\zeta}_0) = \chi^2(\vec{\theta},\, \hat{\vec{\omega}},\,\hat{a};\,\vec{\zeta}_0) - \chi^2(\hat{\hat{\vec{\theta}}},\, \hat{\hat{\vec{\omega}}},\,\hat{\hat{a}};\,\vec{\zeta}_0)\,.
    \label{eq:chi2-prof}
\end{equation}
As explained in \cref{sec:spallation}, we make use of Asimov data sets throughout our analyses. This means that our `observed' data are set to the theoretically expected number of events for each given benchmark point. This leads to two simplifications. Firstly, as the data are perfectly consistent with a given BP, we know that the value of the overall minimised $\chi^2$ will be zero. Secondly, the minimisation over $a$ can be done without resorting to numerical methods for any given $\vec{\theta}$ and $\hat{\vec{\omega}}$. By simply finding that value of $a$ for which $\partial_a(\Delta \chi^2) = 0$, we get the analytical result
\begin{equation}
 \hat{a}=\left[\sum_i\frac{\left(N^i_\text{obs} - N^i_\text{th}\right) N^i_\text{th}}{\left(\sigma^i_\text{stat}\right)^2}\right] \Bigg/ \left[\left(\sigma_\text{sys}\right)^{-2}+\sum_i\left(\frac{N^i_\text{th}}{\sigma^i_\text{stat}}\right)^2\right]\,.
\end{equation}
Note that, since $N_\mathrm{th}^i$ is not a function of $a$, the minimisation over $a$ and $\vec{\omega}$ can be done separately.

Finally, when drawing our contours for the $95\%$ CL regions, we use the fact that our $\Delta \chi^2$ should be distributed according to a $\chi^2$ distribution with $2$ degrees of freedom. This is because, of the $7$ parameters that \cref{eq:chi2-ss} depends on, we profile over $3$ of them in \cref{eq:chi2-prof}, keeping the remaining $2$, represented by $\vec{\zeta}_0$, fixed throughout. We therefore draw the boundaries of our regions at $\Delta \chi^2 = 6.18$.

\subsection{Direct Detection Experiments}
\label{subsec:app_direct}

For DD experiments, we also perform a binned statistical treatment. However, unlike for SS experiments, we assume that the number of counts in each bin follows a Poisson distribution due to the lower number of events expected within the high-energy bins.
Inserting a Poisson likelihood for $\mathcal{L}$ in \cref{eq:llr} and once again exploiting our use of Asimov data sets, we get that
\begin{equation}
\begin{split}
  q(\vec{\theta};\,\vec{\zeta}_0) &= 2\left[\sum_{i = 1}^{N_\mathrm{bins}} (1 + \hat{a}) N_\mathrm{th}^i(\vec{\theta,\,\hat{\omega};\,\vec{\zeta}_0}) - N_\mathrm{obs}^i + N_\mathrm{obs}^i \ln \frac{N_\mathrm{obs}^i}{(1 + \hat{a}) N_\mathrm{th}^i(\vec{\theta,\,\hat{\omega};\,\vec{\zeta}_0}) }\right] + \left(\frac{\hat{a}}{\sigma_\mathrm{^8B}}\right)^2\,.
  \end{split}
  \label{eq:q-dd}   
\end{equation}
Note that, as for SS experiments, we have also introduced the pull parameter $a$ to capture the effect of systematic uncertainties. In the case of DD experiments searching for \cevns, we assume that this is dominated by the uncertainty in the $\mathrm{^8B}$ solar neutrino flux, $\sigma_\mathrm{^8B}$, for which we take different values in the main text.

As before, we can derive the analytical form for $\hat{a}$; we do this by solving the equation $\partial_{a} q = 0$. We find that
\begin{equation}
    \hat{a} = \frac{-(1 + N_\mathrm{th}^\mathrm{tot}\sigma_\mathrm{^8B}^{2}) + \sqrt{(1 + N_\mathrm{th}^\mathrm{tot}\sigma_\mathrm{^8B}^{2})^{2} - 4\sigma_\mathrm{^8B}^{2}(N_\mathrm{th}^\mathrm{tot} - N_\mathrm{obs}^\mathrm{tot})}}{2},\,
    \label{eq:a-dd}
\end{equation}
where $N_\mathrm{obs}^\mathrm{tot}$ and $N_\mathrm{th}^\mathrm{tot}$ are the total observed and theoretically expected number of events across all bins, respectively. We note that in \cref{eq:q-dd,eq:a-dd} we have neglected any background contribution, as the background ($\mathcal{O} (1)$) in DARWIN is expected to be much smaller than the expected signal ($\mathcal{O} (10^{2-3})$) for the majority of bins. Since the pull parameter $a$ only impacts the signal, the analytical minimisation presented in \cref{eq:a-dd} is only possible with zero (or, more generally, constant) background. With a bin-variable background contribution, the minimisation must instead be done numerically.

To draw our $95\%$ CL limits, we make use of Wilks' theorem \cite{Cowan:2010js}. This tells us that the log-likelihood-ratio test statistic asymptotically follows a $\chi^2$ distribution with number of degrees of freedom equal to the difference in the number of free parameters between the null and alternative hypotheses. As previously, this gives us two degrees of freedom. We therefore draw the boundaries of our regions at $q = 6.18$.

\bibliographystyle{JHEP-cerdeno}
\bibliography{references}

\providecommand{\href}[2]{#2}\begingroup\raggedright\begin{thebibliography}{10}

\bibitem{Britton:1992xv}
D.~Britton et~al., \emph{{Improved search for massive neutrinos in $\pi^+\to
  e^+\nu$ decay}},
  \href{http://dx.doi.org/10.1103/PhysRevD.46.R885}{\emph{Phys. Rev. D} {\bf
  46} (1992) 885--887}.

\bibitem{Britton:1993cj}
D.~Britton et~al., \emph{{Measurement of the $\pi^{+} \to e^{+}$ neutrino
  branching ratio}},
  \href{http://dx.doi.org/10.1103/PhysRevD.49.28}{\emph{Phys. Rev. D} {\bf 49}
  (1994) 28--39}.

\bibitem{Aguilar-Arevalo:2019owf}
{\scshape PIENU} collaboration, A.~Aguilar-Arevalo et~al., \emph{{Search for
  heavy neutrinos in $\pi \to \mu\nu$ decay}},
  \href{http://dx.doi.org/10.1016/j.physletb.2019.134980}{\emph{Phys. Lett. B}
  {\bf 798} (2019) 134980}, [\href{http://arxiv.org/abs/1904.03269}{{\tt
  1904.03269}}].

\bibitem{NA62:2020mcv}
{\scshape NA62} collaboration, E.~Cortina~Gil et~al., \emph{{Search for heavy
  neutral lepton production in $K^+$ decays to positrons}},
  \href{http://arxiv.org/abs/2005.09575}{{\tt 2005.09575}}.

\bibitem{T2K:2019jwa}
{\scshape T2K} collaboration, K.~Abe et~al., \emph{{Search for heavy neutrinos
  with the T2K near detector ND280}},
  \href{http://dx.doi.org/10.1103/PhysRevD.100.052006}{\emph{Phys. Rev. D} {\bf
  100} (2019) 052006}, [\href{http://arxiv.org/abs/1902.07598}{{\tt
  1902.07598}}].

\bibitem{Bergsma:1985is}
{\scshape CHARM} collaboration, F.~Bergsma et~al., \emph{{A Search for Decays
  of Heavy Neutrinos in the Mass Range 0.5-{GeV} to 2.8-{GeV}}},
  \href{http://dx.doi.org/10.1016/0370-2693(86)91601-1}{\emph{Phys. Lett. B}
  {\bf 166} (1986) 473--478}.

\bibitem{CooperSarkar:1985nh}
{\scshape WA66} collaboration, A.~M. Cooper-Sarkar et~al., \emph{{Search for
  Heavy Neutrino Decays in the {BEBC} Beam Dump Experiment}},
  \href{http://dx.doi.org/10.1016/0370-2693(85)91493-5}{\emph{Phys. Lett. B}
  {\bf 160} (1985) 207--211}.

\bibitem{Abreu:1996pa}
{\scshape DELPHI} collaboration, P.~Abreu et~al., \emph{{Search for neutral
  heavy leptons produced in Z decays}},
  \href{http://dx.doi.org/10.1007/s002880050370}{\emph{Z. Phys. C} {\bf 74}
  (1997) 57--71}. [Erratum: Z.Phys.C 75, 580 (1997)].

\bibitem{Vaitaitis:1999wq}
{\scshape NuTeV, E815} collaboration, A.~Vaitaitis et~al., \emph{{Search for
  neutral heavy leptons in a high-energy neutrino beam}},
  \href{http://dx.doi.org/10.1103/PhysRevLett.83.4943}{\emph{Phys. Rev. Lett.}
  {\bf 83} (1999) 4943--4946}, [\href{http://arxiv.org/abs/hep-ex/9908011}{{\tt
  hep-ex/9908011}}].

\bibitem{Goldhagen:2021kxe}
K.~Goldhagen, M.~Maltoni, S.~E. Reichard and T.~Schwetz, \emph{{Testing sterile
  neutrino mixing with present and future solar neutrino data}},
  \href{http://dx.doi.org/10.1140/epjc/s10052-022-10052-2}{\emph{Eur. Phys. J.
  C} {\bf 82} (2022) 116}, [\href{http://arxiv.org/abs/2109.14898}{{\tt
  2109.14898}}].

\bibitem{Dentler:2018sju}
M.~Dentler, A.~Hern\'andez-Cabezudo, J.~Kopp, P.~A.~N. Machado, M.~Maltoni,
  I.~Martinez-Soler et~al., \emph{{Updated Global Analysis of Neutrino
  Oscillations in the Presence of eV-Scale Sterile Neutrinos}},
  \href{http://dx.doi.org/10.1007/JHEP08(2018)010}{\emph{JHEP} {\bf 08} (2018)
  010}, [\href{http://arxiv.org/abs/1803.10661}{{\tt 1803.10661}}].

\bibitem{Forero:2021azc}
D.~V. Forero, C.~Giunti, C.~A. Ternes and M.~Tortola, \emph{{Nonunitary
  neutrino mixing in short and long-baseline experiments}},
  \href{http://dx.doi.org/10.1103/PhysRevD.104.075030}{\emph{Phys. Rev. D} {\bf
  104} (2021) 075030}, [\href{http://arxiv.org/abs/2103.01998}{{\tt
  2103.01998}}].

\bibitem{MINOS:2017cae}
{\scshape MINOS+} collaboration, P.~Adamson et~al., \emph{{Search for sterile
  neutrinos in MINOS and MINOS+ using a two-detector fit}},
  \href{http://dx.doi.org/10.1103/PhysRevLett.122.091803}{\emph{Phys. Rev.
  Lett.} {\bf 122} (2019) 091803}, [\href{http://arxiv.org/abs/1710.06488}{{\tt
  1710.06488}}].

\bibitem{Abazajian:2012ys}
K.~N. Abazajian et~al., \emph{{Light Sterile Neutrinos: A White Paper}},
  \href{http://arxiv.org/abs/1204.5379}{{\tt 1204.5379}}.

\bibitem{Sabti:2020yrt}
N.~Sabti, A.~Magalich and A.~Filimonova, \emph{{An Extended Analysis of Heavy
  Neutral Leptons during Big Bang Nucleosynthesis}},
  \href{http://dx.doi.org/10.1088/1475-7516/2020/11/056}{\emph{JCAP} {\bf 11}
  (2020) 056}, [\href{http://arxiv.org/abs/2006.07387}{{\tt 2006.07387}}].

\bibitem{Abdullahi:2022jlv}
A.~M. Abdullahi et~al., \emph{{The present and future status of heavy neutral
  leptons}}, \href{http://dx.doi.org/10.1088/1361-6471/ac98f9}{\emph{J. Phys.
  G} {\bf 50} (2023) 020501}, [\href{http://arxiv.org/abs/2203.08039}{{\tt
  2203.08039}}].

\bibitem{Vincent:2014rja}
A.~C. Vincent, E.~F. Martinez, P.~Hern\'andez, M.~Lattanzi and O.~Mena,
  \emph{{Revisiting cosmological bounds on sterile neutrinos}},
  \href{http://dx.doi.org/10.1088/1475-7516/2015/04/006}{\emph{JCAP} {\bf 04}
  (2015) 006}, [\href{http://arxiv.org/abs/1408.1956}{{\tt 1408.1956}}].

\bibitem{Boyarsky:2021yoh}
A.~Boyarsky, M.~Ovchynnikov, N.~Sabti and V.~Syvolap, \emph{{When feebly
  interacting massive particles decay into neutrinos: The Neff story}},
  \href{http://dx.doi.org/10.1103/PhysRevD.104.035006}{\emph{Phys. Rev. D} {\bf
  104} (2021) 035006}, [\href{http://arxiv.org/abs/2103.09831}{{\tt
  2103.09831}}].

\bibitem{COHERENT:2015mry}
{\scshape COHERENT} collaboration, D.~Akimov et~al., \emph{{The COHERENT
  Experiment at the Spallation Neutron Source}},
  \href{http://arxiv.org/abs/1509.08702}{{\tt 1509.08702}}.

\bibitem{COHERENT:2017ipa}
{\scshape COHERENT} collaboration, D.~Akimov et~al., \emph{{Observation of
  Coherent Elastic Neutrino-Nucleus Scattering}},
  \href{http://dx.doi.org/10.1126/science.aao0990}{\emph{Science} {\bf 357}
  (2017) 1123--1126}, [\href{http://arxiv.org/abs/1708.01294}{{\tt
  1708.01294}}].

\bibitem{COHERENT:2020iec}
{\scshape COHERENT} collaboration, D.~Akimov et~al., \emph{{First Measurement
  of Coherent Elastic Neutrino-Nucleus Scattering on Argon}},
  \href{http://dx.doi.org/10.1103/PhysRevLett.126.012002}{\emph{Phys. Rev.
  Lett.} {\bf 126} (2021) 012002}, [\href{http://arxiv.org/abs/2003.10630}{{\tt
  2003.10630}}].

\bibitem{Freedman:1973yd}
D.~Z. Freedman, \emph{{Coherent Neutrino Nucleus Scattering as a Probe of the
  Weak Neutral Current}},
  \href{http://dx.doi.org/10.1103/PhysRevD.9.1389}{\emph{Phys. Rev. D} {\bf 9}
  (1974) 1389--1392}.

\bibitem{Drukier:1984vhf}
A.~Drukier and L.~Stodolsky, \emph{{Principles and Applications of a Neutral
  Current Detector for Neutrino Physics and Astronomy}},
  \href{http://dx.doi.org/10.1103/PhysRevD.30.2295}{\emph{Phys. Rev. D} {\bf
  30} (1984) 2295}.

\bibitem{Kosmas:2017zbh}
T.~S. Kosmas, D.~K. Papoulias, M.~Tortola and J.~W.~F. Valle, \emph{{Probing
  light sterile neutrino signatures at reactor and Spallation Neutron Source
  neutrino experiments}},
  \href{http://dx.doi.org/10.1103/PhysRevD.96.063013}{\emph{Phys. Rev. D} {\bf
  96} (2017) 063013}, [\href{http://arxiv.org/abs/1703.00054}{{\tt
  1703.00054}}].

\bibitem{Miranda:2020syh}
O.~G. Miranda, D.~K. Papoulias, O.~Sanders, M.~T\'ortola and J.~W.~F. Valle,
  \emph{{Future CEvNS experiments as probes of lepton unitarity and
  light-sterile neutrinos}},
  \href{http://dx.doi.org/10.1103/PhysRevD.102.113014}{\emph{Phys. Rev. D} {\bf
  102} (2020) 113014}, [\href{http://arxiv.org/abs/2008.02759}{{\tt
  2008.02759}}].

\bibitem{Abdullah:2018ykz}
M.~Abdullah, J.~B. Dent, B.~Dutta, G.~L. Kane, S.~Liao and L.~E. Strigari,
  \emph{{Coherent elastic neutrino nucleus scattering as a probe of a Z'
  through kinetic and mass mixing effects}},
  \href{http://dx.doi.org/10.1103/PhysRevD.98.015005}{\emph{Phys. Rev. D} {\bf
  98} (2018) 015005}, [\href{http://arxiv.org/abs/1803.01224}{{\tt
  1803.01224}}].

\bibitem{Bauer:2018onh}
M.~Bauer, P.~Foldenauer and J.~Jaeckel, \emph{{Hunting All the Hidden
  Photons}}, \href{http://dx.doi.org/10.1007/JHEP07(2018)094}{\emph{JHEP} {\bf
  07} (2018) 094}, [\href{http://arxiv.org/abs/1803.05466}{{\tt 1803.05466}}].

\bibitem{Miranda:2020zji}
O.~G. Miranda, D.~K. Papoulias, M.~T\'ortola and J.~W.~F. Valle, \emph{{Probing
  new neutral gauge bosons with $CE\nu NS$ and neutrino-electron scattering}},
  \href{http://dx.doi.org/10.1103/PhysRevD.101.073005}{\emph{Phys. Rev. D} {\bf
  101} (2020) 073005}, [\href{http://arxiv.org/abs/2002.01482}{{\tt
  2002.01482}}].

\bibitem{Bauer:2020itv}
M.~Bauer, P.~Foldenauer and M.~Mosny, \emph{{Flavor structure of anomaly-free
  hidden photon models}},
  \href{http://dx.doi.org/10.1103/PhysRevD.103.075024}{\emph{Phys. Rev. D} {\bf
  103} (2021) 075024}, [\href{http://arxiv.org/abs/2011.12973}{{\tt
  2011.12973}}].

\bibitem{Akimov:2020pdx}
{\scshape COHERENT} collaboration, D.~Akimov et~al., \emph{{First Measurement
  of Coherent Elastic Neutrino-Nucleus Scattering on Argon}},
  \href{http://dx.doi.org/10.1103/PhysRevLett.126.012002}{\emph{Phys. Rev.
  Lett.} {\bf 126} (2021) 012002}, [\href{http://arxiv.org/abs/2003.10630}{{\tt
  2003.10630}}].

\bibitem{COHERENT:2019kwz}
{\scshape COHERENT} collaboration, D.~Akimov et~al., \emph{{Sensitivity of the
  COHERENT Experiment to Accelerator-Produced Dark Matter}},
  \href{http://dx.doi.org/10.1103/PhysRevD.102.052007}{\emph{Phys. Rev. D} {\bf
  102} (2020) 052007}, [\href{http://arxiv.org/abs/1911.06422}{{\tt
  1911.06422}}].

\bibitem{ccm}
``Coherent captain mills.'' \url{https://p25ext.lanl.gov/%7Elee/CaptainMills/}.

\bibitem{Baxter:2019mcx}
D.~Baxter et~al., \emph{{Coherent Elastic Neutrino-Nucleus Scattering at the
  European Spallation Source}},
  \href{http://dx.doi.org/10.1007/JHEP02(2020)123}{\emph{JHEP} {\bf 02} (2020)
  123}, [\href{http://arxiv.org/abs/1911.00762}{{\tt 1911.00762}}].

\bibitem{Miranda:2020tif}
O.~G. Miranda, D.~K. Papoulias, G.~Sanchez~Garcia, O.~Sanders, M.~T\'ortola and
  J.~W.~F. Valle, \emph{{Implications of the first detection of coherent
  elastic neutrino-nucleus scattering (CEvNS) with Liquid Argon}},
  \href{http://dx.doi.org/10.1007/JHEP05(2020)130}{\emph{JHEP} {\bf 05} (2020)
  130}, [\href{http://arxiv.org/abs/2003.12050}{{\tt 2003.12050}}]. [Erratum:
  JHEP 01, 067 (2021)].

\bibitem{Banerjee:2018eaf}
H.~Banerjee, P.~Byakti and S.~Roy, \emph{{Supersymmetric gauged
  U(1)$_{L_{\mu}-L_{\tau}}$ model for neutrinos and the muon $(g-2)$ anomaly}},
  \href{http://dx.doi.org/10.1103/PhysRevD.98.075022}{\emph{Phys. Rev. D} {\bf
  98} (2018) 075022}, [\href{http://arxiv.org/abs/1805.04415}{{\tt
  1805.04415}}].

\bibitem{Papoulias:2019txv}
D.~K. Papoulias, \emph{{COHERENT constraints after the COHERENT-2020 quenching
  factor measurement}},
  \href{http://dx.doi.org/10.1103/PhysRevD.102.113004}{\emph{Phys. Rev. D} {\bf
  102} (2020) 113004}, [\href{http://arxiv.org/abs/1907.11644}{{\tt
  1907.11644}}].

\bibitem{Khan:2019cvi}
A.~N. Khan and W.~Rodejohann, \emph{{New physics from COHERENT data with an
  improved quenching factor}},
  \href{http://dx.doi.org/10.1103/PhysRevD.100.113003}{\emph{Phys. Rev. D} {\bf
  100} (2019) 113003}, [\href{http://arxiv.org/abs/1907.12444}{{\tt
  1907.12444}}].

\bibitem{Bolton:2021pey}
P.~D. Bolton, F.~F. Deppisch, K.~Fridell, J.~Harz, C.~Hati and S.~Kulkarni,
  \emph{{Probing active-sterile neutrino transition magnetic moments with
  photon emission from CE\ensuremath{\nu}NS}},
  \href{http://dx.doi.org/10.1103/PhysRevD.106.035036}{\emph{Phys. Rev. D} {\bf
  106} (2022) 035036}, [\href{http://arxiv.org/abs/2110.02233}{{\tt
  2110.02233}}].

\bibitem{Miranda:2021kre}
O.~G. Miranda, D.~K. Papoulias, O.~Sanders, M.~T\'ortola and J.~W.~F. Valle,
  \emph{{Low-energy probes of sterile neutrino transition magnetic moments}},
  \href{http://dx.doi.org/10.1007/JHEP12(2021)191}{\emph{JHEP} {\bf 12} (2021)
  191}, [\href{http://arxiv.org/abs/2109.09545}{{\tt 2109.09545}}].

\bibitem{DeRomeri:2022twg}
V.~De~Romeri, O.~G. Miranda, D.~K. Papoulias, G.~Sanchez~Garcia, M.~T\'ortola
  and J.~W.~F. Valle, \emph{{Physics implications of a combined analysis of
  COHERENT CsI and LAr data}},  \href{http://arxiv.org/abs/2211.11905}{{\tt
  2211.11905}}.

\bibitem{Candela:2023rvt}
P.~M. Candela, V.~De~Romeri and D.~K. Papoulias, \emph{{COHERENT production of
  a Dark Fermion}},  \href{http://arxiv.org/abs/2305.03341}{{\tt 2305.03341}}.

\bibitem{Cerdeno:2016sfi}
D.~G. Cerdeño, M.~Fairbairn, T.~Jubb, P.~A.~N. Machado, A.~C. Vincent and
  C.~Bœhm, \emph{{Physics from solar neutrinos in dark matter direct detection
  experiments}}, \href{http://dx.doi.org/10.1007/JHEP09(2016)048,
  10.1007/JHEP05(2016)118}{\emph{JHEP} {\bf 05} (2016) 118},
  [\href{http://arxiv.org/abs/1604.01025}{{\tt 1604.01025}}]. [Erratum:
  JHEP09,048(2016)].

\bibitem{Dutta:2017nht}
B.~Dutta, S.~Liao, L.~E. Strigari and J.~W. Walker, \emph{{Non-standard
  interactions of solar neutrinos in dark matter experiments}},
  \href{http://dx.doi.org/10.1016/j.physletb.2017.08.031}{\emph{Phys. Lett. B}
  {\bf 773} (2017) 242--246}, [\href{http://arxiv.org/abs/1705.00661}{{\tt
  1705.00661}}].

\bibitem{Gelmini:2018gqa}
G.~B. Gelmini, V.~Takhistov and S.~J. Witte, \emph{{Geoneutrinos in Large
  Direct Detection Experiments}},
  \href{http://dx.doi.org/10.1103/PhysRevD.99.093009}{\emph{Phys. Rev. D} {\bf
  99} (2019) 093009}, [\href{http://arxiv.org/abs/1812.05550}{{\tt
  1812.05550}}].

\bibitem{Essig:2018tss}
R.~Essig, M.~Sholapurkar and T.-T. Yu, \emph{{Solar Neutrinos as a Signal and
  Background in Direct-Detection Experiments Searching for Sub-GeV Dark Matter
  With Electron Recoils}},
  \href{http://dx.doi.org/10.1103/PhysRevD.97.095029}{\emph{Phys. Rev. D} {\bf
  97} (2018) 095029}, [\href{http://arxiv.org/abs/1801.10159}{{\tt
  1801.10159}}].

\bibitem{Amaral:2020tga}
D.~W. P.~d. Amaral, D.~G. Cerdeno, P.~Foldenauer and E.~Reid, \emph{{Solar
  neutrino probes of the muon anomalous magnetic moment in the gauged $
  \mathrm{U}{(1)}_{L_{\mu }-{L}_{\tau }} $}},
  \href{http://dx.doi.org/10.1007/JHEP12(2020)155}{\emph{JHEP} {\bf 12} (2020)
  155}, [\href{http://arxiv.org/abs/2006.11225}{{\tt 2006.11225}}].

\bibitem{Dutta:2020che}
B.~Dutta, R.~F. Lang, S.~Liao, S.~Sinha, L.~Strigari and A.~Thompson, \emph{{A
  global analysis strategy to resolve neutrino NSI degeneracies with scattering
  and oscillation data}},
  \href{http://dx.doi.org/10.1007/JHEP09(2020)106}{\emph{JHEP} {\bf 09} (2020)
  106}, [\href{http://arxiv.org/abs/2002.03066}{{\tt 2002.03066}}].

\bibitem{Amaral:2021rzw}
D.~W.~P. Amaral, D.~G. Cerdeno, A.~Cheek and P.~Foldenauer, \emph{{Confirming
  $U(1)_{L_\mu -L_{\tau }}$ as a solution for $(g-2)_\mu $ with neutrinos}},
  \href{http://dx.doi.org/10.1140/epjc/s10052-021-09670-z}{\emph{Eur. Phys. J.
  C} {\bf 81} (2021) 861}, [\href{http://arxiv.org/abs/2104.03297}{{\tt
  2104.03297}}].

\bibitem{Munoz:2021sad}
V.~Munoz, V.~Takhistov, S.~J. Witte and G.~M. Fuller, \emph{{Exploring the
  origin of supermassive black holes with coherent neutrino scattering}},
  \href{http://dx.doi.org/10.1088/1475-7516/2021/11/020}{\emph{JCAP} {\bf 11}
  (2021) 020}, [\href{http://arxiv.org/abs/2102.00885}{{\tt 2102.00885}}].

\bibitem{deGouvea:2021ymm}
A.~de~Gouv\^ea, E.~McGinness, I.~Martinez-Soler and Y.~F. Perez-Gonzalez,
  \emph{{pp solar neutrinos at DARWIN}},
  \href{http://dx.doi.org/10.1103/PhysRevD.106.096017}{\emph{Phys. Rev. D} {\bf
  106} (2022) 096017}, [\href{http://arxiv.org/abs/2111.02421}{{\tt
  2111.02421}}].

\bibitem{Amaral:2023tbs}
D.~W.~P. Amaral, D.~Cerdeno, A.~Cheek and P.~Foldenauer, \emph{{A direct
  detection view of the neutrino NSI landscape}},
  \href{http://arxiv.org/abs/2302.12846}{{\tt 2302.12846}}.

\bibitem{Shoemaker:2018vii}
I.~M. Shoemaker and J.~Wyenberg, \emph{{Direct Detection Experiments at the
  Neutrino Dipole Portal Frontier}},
  \href{http://dx.doi.org/10.1103/PhysRevD.99.075010}{\emph{Phys. Rev. D} {\bf
  99} (2019) 075010}, [\href{http://arxiv.org/abs/1811.12435}{{\tt
  1811.12435}}].

\bibitem{XENON:2020rca}
{\scshape XENON} collaboration, E.~Aprile et~al., \emph{{Excess electronic
  recoil events in XENON1T}},
  \href{http://dx.doi.org/10.1103/PhysRevD.102.072004}{\emph{Phys. Rev. D} {\bf
  102} (2020) 072004}, [\href{http://arxiv.org/abs/2006.09721}{{\tt
  2006.09721}}].

\bibitem{Shoemaker:2020kji}
I.~M. Shoemaker, Y.-D. Tsai and J.~Wyenberg, \emph{{Active-to-sterile neutrino
  dipole portal and the XENON1T excess}},
  \href{http://dx.doi.org/10.1103/PhysRevD.104.115026}{\emph{Phys. Rev. D} {\bf
  104} (2021) 115026}, [\href{http://arxiv.org/abs/2007.05513}{{\tt
  2007.05513}}].

\bibitem{Brdar:2020quo}
V.~Brdar, A.~Greljo, J.~Kopp and T.~Opferkuch, \emph{{The Neutrino Magnetic
  Moment Portal: Cosmology, Astrophysics, and Direct Detection}},
  \href{http://dx.doi.org/10.1088/1475-7516/2021/01/039}{\emph{JCAP} {\bf 01}
  (2021) 039}, [\href{http://arxiv.org/abs/2007.15563}{{\tt 2007.15563}}].

\bibitem{XENON:2022ltv}
{\scshape XENON} collaboration, E.~Aprile et~al., \emph{{Search for New Physics
  in Electronic Recoil Data from XENONnT}},
  \href{http://dx.doi.org/10.1103/PhysRevLett.129.161805}{\emph{Phys. Rev.
  Lett.} {\bf 129} (2022) 161805}, [\href{http://arxiv.org/abs/2207.11330}{{\tt
  2207.11330}}].

\bibitem{Pospelov:2011ha}
M.~Pospelov, \emph{{Neutrino Physics with Dark Matter Experiments and the
  Signature of New Baryonic Neutral Currents}},
  \href{http://dx.doi.org/10.1103/PhysRevD.84.085008}{\emph{Phys. Rev. D} {\bf
  84} (2011) 085008}, [\href{http://arxiv.org/abs/1103.3261}{{\tt 1103.3261}}].

\bibitem{Arguelles:2022lzs}
C.~A. Arg\"uelles, N.~Foppiani and M.~Hostert, \emph{{Efficiently exploring
  multidimensional parameter spaces beyond the Standard Model}},
  \href{http://dx.doi.org/10.1103/PhysRevD.107.035027}{\emph{Phys. Rev. D} {\bf
  107} (2023) 035027}, [\href{http://arxiv.org/abs/2205.12273}{{\tt
  2205.12273}}].

\bibitem{Helm:1956zz}
R.~H. Helm, \emph{{Inelastic and Elastic Scattering of 187-Mev Electrons from
  Selected Even-Even Nuclei}},
  \href{http://dx.doi.org/10.1103/PhysRev.104.1466}{\emph{Phys. Rev.} {\bf 104}
  (1956) 1466--1475}.

\bibitem{Lewin:1995rx}
J.~D. Lewin and P.~F. Smith, \emph{{Review of mathematics, numerical factors,
  and corrections for dark matter experiments based on elastic nuclear
  recoil}},
  \href{http://dx.doi.org/10.1016/S0927-6505(96)00047-3}{\emph{Astropart.
  Phys.} {\bf 6} (1996) 87--112}.

\bibitem{Foguel:2022ppx}
A.~L. Foguel, P.~Reimitz and R.~Z. Funchal, \emph{{A robust description of
  hadronic decays in light vector mediator models}},
  \href{http://dx.doi.org/10.1007/JHEP04(2022)119}{\emph{JHEP} {\bf 04} (2022)
  119}, [\href{http://arxiv.org/abs/2201.01788}{{\tt 2201.01788}}].

\bibitem{Coloma:2017egw}
P.~Coloma, P.~B. Denton, M.~C. Gonzalez-Garcia, M.~Maltoni and T.~Schwetz,
  \emph{{Curtailing the Dark Side in Non-Standard Neutrino Interactions}},
  \href{http://dx.doi.org/10.1007/JHEP04(2017)116}{\emph{JHEP} {\bf 04} (2017)
  116}, [\href{http://arxiv.org/abs/1701.04828}{{\tt 1701.04828}}].

\bibitem{Wolfenstein:1977ue}
L.~Wolfenstein, \emph{{Neutrino Oscillations in Matter}},
  \href{http://dx.doi.org/10.1103/PhysRevD.17.2369}{\emph{Phys. Rev. D} {\bf
  17} (1978) 2369--2374}.

\bibitem{Guzzo:1991cp}
M.~M. Guzzo and S.~T. Petcov, \emph{{On the matter enhanced transitions of
  solar neutrinos in the absence of neutrino mixing in vacuum}},
  \href{http://dx.doi.org/10.1016/0370-2693(91)91295-7}{\emph{Phys. Lett. B}
  {\bf 271} (1991) 172--178}.

\bibitem{Guzzo:1991hi}
M.~M. Guzzo, A.~Masiero and S.~T. Petcov, \emph{{On the MSW effect with
  massless neutrinos and no mixing in the vacuum}},
  \href{http://dx.doi.org/10.1016/0370-2693(91)90984-X}{\emph{Phys. Lett. B}
  {\bf 260} (1991) 154--160}.

\bibitem{Gonzalez-Garcia:1998ryc}
M.~C. Gonzalez-Garcia, M.~M. Guzzo, P.~I. Krastev, H.~Nunokawa, O.~L.~G. Peres,
  V.~Pleitez et~al., \emph{{Atmospheric neutrino observations and flavor
  changing interactions}},
  \href{http://dx.doi.org/10.1103/PhysRevLett.82.3202}{\emph{Phys. Rev. Lett.}
  {\bf 82} (1999) 3202--3205}, [\href{http://arxiv.org/abs/hep-ph/9809531}{{\tt
  hep-ph/9809531}}].

\bibitem{Bergmann:2000gp}
S.~Bergmann, M.~M. Guzzo, P.~C. de~Holanda, P.~I. Krastev and H.~Nunokawa,
  \emph{{Status of the solution to the solar neutrino problem based on
  nonstandard neutrino interactions}},
  \href{http://dx.doi.org/10.1103/PhysRevD.62.073001}{\emph{Phys. Rev. D} {\bf
  62} (2000) 073001}, [\href{http://arxiv.org/abs/hep-ph/0004049}{{\tt
  hep-ph/0004049}}].

\bibitem{Guzzo:2000kx}
M.~M. Guzzo, H.~Nunokawa, P.~C. de~Holanda and O.~L.~G. Peres, \emph{{On the
  massless 'just-so' solution to the solar neutrino problem}},
  \href{http://dx.doi.org/10.1103/PhysRevD.64.097301}{\emph{Phys. Rev. D} {\bf
  64} (2001) 097301}, [\href{http://arxiv.org/abs/hep-ph/0012089}{{\tt
  hep-ph/0012089}}].

\bibitem{Guzzo:2001mi}
M.~Guzzo, P.~C. de~Holanda, M.~Maltoni, H.~Nunokawa, M.~A. Tortola and J.~W.~F.
  Valle, \emph{{Status of a hybrid three neutrino interpretation of neutrino
  data}}, \href{http://dx.doi.org/10.1016/S0550-3213(02)00139-6}{\emph{Nucl.
  Phys. B} {\bf 629} (2002) 479--490},
  [\href{http://arxiv.org/abs/hep-ph/0112310}{{\tt hep-ph/0112310}}].

\bibitem{GonzalezGarcia:2004wg}
M.~C. Gonzalez-Garcia and M.~Maltoni, \emph{{Atmospheric neutrino oscillations
  and new physics}},
  \href{http://dx.doi.org/10.1103/PhysRevD.70.033010}{\emph{Phys. Rev. D} {\bf
  70} (2004) 033010}, [\href{http://arxiv.org/abs/hep-ph/0404085}{{\tt
  hep-ph/0404085}}].

\bibitem{Arguelles_2023}
C.~A. Argüelles, G.~Barenboim, M.~Bustamante, P.~Coloma, P.~B. Denton,
  I.~Esteban et~al., \emph{Snowmass white paper: beyond the standard model
  effects on neutrino flavor},
  \href{http://dx.doi.org/10.1140/epjc/s10052-022-11049-7}{\emph{The European
  Physical Journal C} {\bf 83} (jan, 2023) }.

\bibitem{Cowan:2010js}
G.~Cowan, K.~Cranmer, E.~Gross and O.~Vitells, \emph{{Asymptotic formulae for
  likelihood-based tests of new physics}},
  \href{http://dx.doi.org/10.1140/epjc/s10052-011-1554-0}{\emph{Eur. Phys. J.
  C} {\bf 71} (2011) 1554}, [\href{http://arxiv.org/abs/1007.1727}{{\tt
  1007.1727}}]. [Erratum: Eur.Phys.J.C 73, 2501 (2013)].

\bibitem{SNuDD-code}
D.~W.~P. Amaral, D.~G. Cerde\~no, A.~Cheek and P.~Foldenauer,
  \emph{\textnormal{SNuDD [Computer Software]}, available at
  \href{https://github.com/SNuDD/SNuDD.git}{\textnormal{https://github.com/snudd/snudd.git}}},
  2023.

\bibitem{multinest}
F.~{Feroz}, M.~P. {Hobson} and M.~{Bridges}, \emph{{MULTINEST: an efficient and
  robust Bayesian inference tool for cosmology and particle physics}},
  \href{http://dx.doi.org/10.1111/j.1365-2966.2009.14548.x}{\emph{\mnras} {\bf
  398} (Oct., 2009) 1601--1614}, [\href{http://arxiv.org/abs/0809.3437}{{\tt
  0809.3437}}].

\bibitem{multinest2}
F.~{Feroz}, M.~P. {Hobson}, E.~{Cameron} and A.~N. {Pettitt}, \emph{{Importance
  Nested Sampling and the MultiNest Algorithm}},
  \href{http://dx.doi.org/10.21105/astro.1306.2144}{\emph{The Open Journal of
  Astrophysics} {\bf 2} (Nov., 2019) 10},
  [\href{http://arxiv.org/abs/1306.2144}{{\tt 1306.2144}}].

\bibitem{pymultinest}
J.~{Buchner}, A.~{Georgakakis}, K.~{Nandra}, L.~{Hsu}, C.~{Rangel},
  M.~{Brightman} et~al., \emph{{X-ray spectral modelling of the AGN obscuring
  region in the CDFS: Bayesian model selection and catalogue}},
  \href{http://dx.doi.org/10.1051/0004-6361/201322971}{\emph{\aap} {\bf 564}
  (Apr., 2014) A125}, [\href{http://arxiv.org/abs/1402.0004}{{\tt 1402.0004}}].

\bibitem{COHERENT:2021xmm}
{\scshape COHERENT} collaboration, D.~Akimov et~al., \emph{{Measurement of the
  Coherent Elastic Neutrino-Nucleus Scattering Cross Section on CsI by
  COHERENT}},
  \href{http://dx.doi.org/10.1103/PhysRevLett.129.081801}{\emph{Phys. Rev.
  Lett.} {\bf 129} (2022) 081801}, [\href{http://arxiv.org/abs/2110.07730}{{\tt
  2110.07730}}].

\bibitem{LZ:2018qzl}
{\scshape LZ} collaboration, D.~S. Akerib et~al., \emph{{Projected WIMP
  sensitivity of the LUX-ZEPLIN dark matter experiment}},
  \href{http://dx.doi.org/10.1103/PhysRevD.101.052002}{\emph{Phys. Rev. D} {\bf
  101} (2020) 052002}, [\href{http://arxiv.org/abs/1802.06039}{{\tt
  1802.06039}}].

\bibitem{XENON:2020kmp}
{\scshape XENON} collaboration, E.~Aprile et~al., \emph{{Projected WIMP
  sensitivity of the XENONnT dark matter experiment}},
  \href{http://dx.doi.org/10.1088/1475-7516/2020/11/031}{\emph{JCAP} {\bf 11}
  (2020) 031}, [\href{http://arxiv.org/abs/2007.08796}{{\tt 2007.08796}}].

\bibitem{DARWIN:2016hyl}
{\scshape DARWIN} collaboration, J.~Aalbers et~al., \emph{{DARWIN: towards the
  ultimate dark matter detector}},
  \href{http://dx.doi.org/10.1088/1475-7516/2016/11/017}{\emph{JCAP} {\bf 11}
  (2016) 017}, [\href{http://arxiv.org/abs/1606.07001}{{\tt 1606.07001}}].

\bibitem{OHare:2021utq}
C.~A.~J. O'Hare, \emph{{New Definition of the Neutrino Floor for Direct Dark
  Matter Searches}},
  \href{http://dx.doi.org/10.1103/PhysRevLett.127.251802}{\emph{Phys. Rev.
  Lett.} {\bf 127} (2021) 251802}, [\href{http://arxiv.org/abs/2109.03116}{{\tt
  2109.03116}}].

\bibitem{Coloma:2022umy}
P.~Coloma, M.~C. Gonzalez-Garcia, M.~Maltoni, J.~P. Pinheiro and S.~Urrea,
  \emph{{Constraining new physics with Borexino Phase-II spectral data}},
  \href{http://dx.doi.org/10.1007/JHEP07(2022)138}{\emph{JHEP} {\bf 07} (2022)
  138}, [\href{http://arxiv.org/abs/2204.03011}{{\tt 2204.03011}}]. [Erratum:
  JHEP 11, 138 (2022)].

\bibitem{Vinyoles:2016djt}
N.~Vinyoles, A.~M. Serenelli, F.~L. Villante, S.~Basu, J.~Bergstr\"om, M.~C.
  Gonzalez-Garcia et~al., \emph{{A new Generation of Standard Solar Models}},
  \href{http://dx.doi.org/10.3847/1538-4357/835/2/202}{\emph{Astrophys. J.}
  {\bf 835} (2017) 202}, [\href{http://arxiv.org/abs/1611.09867}{{\tt
  1611.09867}}].

\bibitem{Esteban:2020cvm}
I.~Esteban, M.~C. Gonzalez-Garcia, M.~Maltoni, T.~Schwetz and A.~Zhou,
  \emph{{The fate of hints: updated global analysis of three-flavor neutrino
  oscillations}}, \href{http://dx.doi.org/10.1007/JHEP09(2020)178}{\emph{JHEP}
  {\bf 09} (2020) 178}, [\href{http://arxiv.org/abs/2007.14792}{{\tt
  2007.14792}}].

\bibitem{Pospelov:2012gm}
M.~Pospelov and J.~Pradler, \emph{{Elastic scattering signals of solar
  neutrinos with enhanced baryonic currents}},
  \href{http://dx.doi.org/10.1103/PhysRevD.85.113016}{\emph{Phys. Rev. D} {\bf
  85} (2012) 113016}, [\href{http://arxiv.org/abs/1203.0545}{{\tt 1203.0545}}].
  [Erratum: Phys.Rev.D 88, 039904 (2013)].

\bibitem{Bahcall:2004pz}
J.~N. Bahcall, A.~M. Serenelli and S.~Basu, \emph{{New solar opacities,
  abundances, helioseismology, and neutrino fluxes}},
  \href{http://dx.doi.org/10.1086/428929}{\emph{Astrophys. J. Lett.} {\bf 621}
  (2005) L85--L88}, [\href{http://arxiv.org/abs/astro-ph/0412440}{{\tt
  astro-ph/0412440}}].

\bibitem{Bergstrom:2016cbh}
J.~Bergstrom, M.~C. Gonzalez-Garcia, M.~Maltoni, C.~Pena-Garay, A.~M. Serenelli
  and N.~Song, \emph{{Updated determination of the solar neutrino fluxes from
  solar neutrino data}},
  \href{http://dx.doi.org/10.1007/JHEP03(2016)132}{\emph{JHEP} {\bf 03} (2016)
  132}, [\href{http://arxiv.org/abs/1601.00972}{{\tt 1601.00972}}].

\bibitem{Capozzi:2018dat}
F.~Capozzi, S.~W. Li, G.~Zhu and J.~F. Beacom, \emph{{DUNE as the
  Next-Generation Solar Neutrino Experiment}},
  \href{http://dx.doi.org/10.1103/PhysRevLett.123.131803}{\emph{Phys. Rev.
  Lett.} {\bf 123} (2019) 131803}, [\href{http://arxiv.org/abs/1808.08232}{{\tt
  1808.08232}}].

\bibitem{SNO:2011hx}
{\scshape SNO} collaboration, B.~Aharmim et~al., \emph{{Combined Analysis of
  all Three Phases of Solar Neutrino Data from the Sudbury Neutrino
  Observatory}},
  \href{http://dx.doi.org/10.1103/PhysRevC.88.025501}{\emph{Phys. Rev. C} {\bf
  88} (2013) 025501}, [\href{http://arxiv.org/abs/1109.0763}{{\tt 1109.0763}}].

\end{thebibliography}\endgroup


\end{document}